\documentclass[a4paper,12pt]{article}

\usepackage{pslatex}
\usepackage{textpos}
\usepackage{caption}
\usepackage{tikz}

\usepackage{algorithm}
\usepackage{algpseudocode}
\usepackage{listings}
\usepackage{color}
\usepackage{textcomp}
\usepackage{graphicx,xcolor}
\usepackage{epsfig}
\usepackage{epstopdf}

\usepackage{pgf}

\usepackage{amsmath,amssymb,amsfonts,amsthm,breqn,amsbsy}
\usepackage{longtable}
\usepackage{verbatim}
\usepackage{float}
\usepackage{appendix}
\usepackage{array}
\usepackage{varwidth}
\usepackage{multirow}
\usepackage{tabularx}
\usepackage{framed}
\usepackage{lipsum}
\usepackage[round]{natbib}

\usepackage[hidelinks]{hyperref}
\usepackage{longtable}

\usepackage[utf8]{inputenc}
\usepackage{booktabs}
\usepackage{lscape}
\usepackage{rotating}
\usepackage{fancyvrb}

\usepackage{setspace}
\usepackage[T1]{fontenc}
\usepackage{indentfirst}
\usepackage{lineno}
\usepackage{mathtools}
\usepackage{bigints}
\usepackage{url}
\usepackage{bm,bbm}
\usepackage[normalem]{ulem}

\def\spacingset#1{\renewcommand{\baselinestretch}%
{#1}\small\normalsize} \spacingset{1}
\def\htheta{\hat{\theta}}

\def\param{\theta}
\def\paramspa{\Theta}

\def\models{\mathcal{M}}

\def\rset{\mathbb{R}}

\def\htheta{\hat{\theta}}

\def\rset{\mathbb{R}}
\DeclareMathOperator*{\argmax}{arg\,max}
\newcommand{\pkg}[1]{\texttt{#1}}

\newcommand{\blind}{0}

\begin{document}

% http://www.stat.columbia.edu/~yangfeng/pubs/2014-YF-JCGS.pdf

%====================================================================================
%====================================================================================
\if0\blind
{
  \title{\bf Bayesian model selection for exponential random graph models via adjusted pseudolikelihoods}
  \author{Lampros Bouranis\thanks{
	The authors thank the editor, the associate editor and the anonymous referees for their constructive comments that helped improve the article. The Insight Centre for Data Analytics is supported by Science Foundation Ireland under Grant Number SFI/12/RC/2289. Nial Friel's research was also supported by a Science Foundation Ireland grant: 12/IP/1424.
		},
    Nial Friel, Florian Maire\hspace{.2cm}\\
    School of Mathematics and Statistics \& Insight Centre for Data Analytics,\\
    University College Dublin, Ireland
		}
  \maketitle
} \fi

\if1\blind
{
  \bigskip
  \bigskip
  \bigskip
  \begin{center}
    {\LARGE\bf Title}
\end{center}
  \medskip
} \fi

\bigskip
%====================================================================================
%====================================================================================
\begin{abstract}
Models with intractable likelihood functions arise in areas including network analysis and spatial statistics, especially those involving Gibbs random fields. Posterior parameter estimation in these settings is termed a doubly-intractable problem because both the likelihood function and the posterior distribution are intractable.
The comparison of Bayesian models is often based on the statistical evidence, the integral of the un-normalised posterior distribution over the model parameters which is rarely available in closed form. For doubly-intractable models, estimating the evidence adds another layer of difficulty.
Consequently, the selection of the model that best describes an observed network among a collection of exponential random graph models for network analysis is a daunting task. Pseudolikelihoods offer a tractable approximation to the likelihood but should be treated with caution because they can lead to an unreasonable inference. This paper specifies a method to adjust pseudolikelihoods in order to obtain a reasonable, yet tractable, approximation to the likelihood. This allows implementation of widely used computational methods for evidence estimation and pursuit of Bayesian model selection of exponential random graph models for the analysis of social networks.
Empirical comparisons to existing methods show that our procedure yields similar evidence estimates, but at a lower computational cost.
\end{abstract}

\noindent%
{\it Keywords:} Bayes factors, Evidence,  Intractable normalising constants.
\vfill

\newpage
\spacingset{1.45} % DON'T change the spacing!

%====================================================================================
%====================================================================================
\section{Introduction}\label{section:intro}
Bayesian inference for models that are characterized by an intractable likelihood function has received considerable attention by the statistical community, notably the class of Gibbs random fields (GRFs). Popular examples include the autologistic model \citep{besag3}, used to model the spatial distribution of binary random variables defined on a lattice or grid and the exponential random graph model (ERGM) for social network analysis \citep{robins2}. Despite their popularity, posterior parameter estimation for GRFs presents considerable difficulties because the normalising constant $z(\param)$ of the likelihood density
%%%%%%%%%%%%%%%%%%%%%%%%%%%%%%%%%%%%%%%%%%
\begin{equation}\label{intro_eqn:likelihood}
f(y\mid\theta)=\frac{q(y\mid\param)}{z(\param)}
\end{equation}
%%%%%%%%%%%%%%%%%%%%%%%%%%%%%%%%%%%%%%%%%%
is typically intractable for all but trivially small graphs. The posterior distribution defined as
%%%%%%%%%%%%%%%%%%%%%%%%%%%%%%%%%%%%%%%%%%
\begin{equation}\label{eq:posterior}
\pi(\param\mid y)=
\frac{f(y\mid \param)p(\param)}{\pi(y)}=\frac{f(y\mid \param)p(\param)}{\int_{\paramspa} f(y\mid \param)p(\param)\;d\param}
\end{equation}
%%%%%%%%%%%%%%%%%%%%%%%%%%%%%%%%%%%%%%%%%%
is termed \textit{doubly-intractable} because of the intractability of the normalising term of the likelihood model within the posterior and the intractability of the posterior normalising term.

Bayesian model comparison is often achieved by estimating the Bayes factor \citep{kassr95}, which relies upon the \textit{marginal likelihood} or model \textit{evidence}, $\pi(y)$, of each of the competing models. However, for many models of interest with intractable likelihoods such as GRFs, estimation of the marginal likelihood adds another layer of difficulty. This paper addresses this problem in the context of Bayesian model comparison of exponential random graph models.

Related work by \cite{friel4} and \cite{johansen} has the same objective as our study, namely to estimate the evidence in the presence of an intractable likelihood normalising constant. Contrary to our method, 
their proposed algorithms rely heavily on repeated simulations from the likelihood. \cite{friel4} devised a "population" version of the exchange algorithm \citep{moller}, however for evidence estimation, it is 
limited to models with a small number of parameters.
\cite{johansen} describe an importance sampling approach for estimating the evidence, which is promising for low-dimensional models. However, when moving to higher dimensional settings their approach makes use of a 
particle filter to estimate the evidence, which is naturally more computationally demanding. 
%Additionally, those approaches rely on simulation to circumvent the evaluation of the intractable likelihood, adding significantly to the computational burden.

Motivated by overcoming the intractability of the likelihood in \eqref{intro_eqn:likelihood}, a natural approach is to use composite likelihoods as a plug-in for the true likelihood \citep{varin1}. The pseudolikelihood \citep{besag1} is an antecedent of composite likelihoods and was developed in the context of ERGMs by \cite{strauss}.
Building on the work of \cite{friel1}, \cite{bouranismisp} proposed an alternative approach to Bayesian inference for ERGMs. The replacement of the true likelihood with the pseudolikelihood approximation in Bayes formula yields what is termed a \textit{pseudo-posterior} distribution, a tractable Bayesian model from which it is straightforward to sample.
Bayesian inference based on the pseudolikelihood can be problematic however, as in some cases the posterior mean estimates are biased and the posterior variances are typically underestimated.
\cite{bouranismisp} developed an approach to allow for correction of a sample from the pseudo-posterior distribution so that it is approximately distributed from the target posterior distribution.
%%%%%%%%%%%%%%%%%%%%%%%%%%%%%%%%%%%%%%%%%%
\begin{figure}[H]
\centering
\includegraphics[width=14cm,height=9cm]{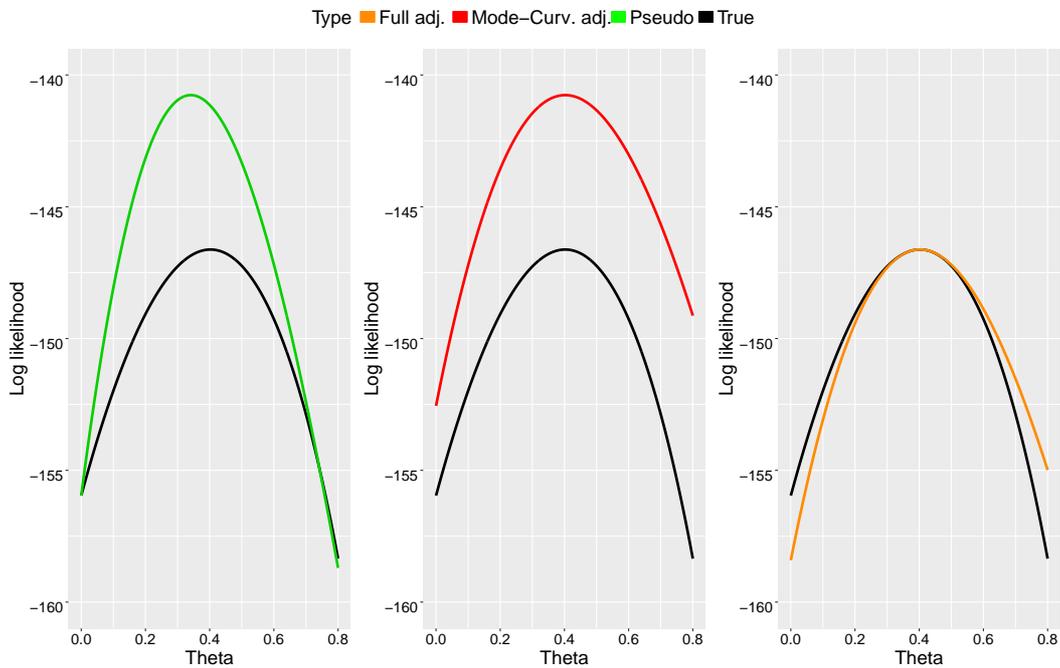}
\caption{A graphical representation of the steps involved in the adjustment of the log-pseudolikelihood. A mode and curvature-adjusted log-pseudolikelihood (red curve) stems from the unadjusted log-pseudolikelihood (green curve). The magnitude adjustment ensures equality with the true log-likelihood (black curve) at the mode.}
\label{fig:correction_graph}
\end{figure}
%%%%%%%%%%%%%%%%%%%%%%%%%%%%%%%%%%%%%%%%%%
While parameter inference based on the adjusted pseudo-posterior distribution yields reasonable results \citep{bouranismisp}, evidence estimation using this correction procedure is inefficient, a point which is 
explained in Section \ref{section:BF_approx}.
Based on this observation, we consider in this paper  adjusting the pseudolikelihood directly, as opposed to the pseudo-posterior, and the likelihood function in the model evidence is replaced with this 
\textit{fully adjusted} pseudolikelihood. These adjustments involve a correction of the mode, the curvature and the magnitude at the mode of the pseudolikelihood function, as outlined in Figure \ref{fig:correction_graph}. 
The crucial point is that this adjusted pseudolikelihood function renders the corresponding posterior distribution amendable to standard evidence estimation methods from the Bayesian toolbox. A non-exhaustive list of 
such methods includes Chib's method \citep{chib1} and its extension \citep{chib2}, importance sampling \citep{liu}, annealed importance sampling \citep{neal}, bridge sampling \citep{meng} and path sampling/thermodynamic integration \citep{gelman,lartillot,friel5,calderhead}, among others. The tractability of the fully adjusted pseudolikelihood allows for evidence estimation with such methods, thereby allowing Bayesian model selection of exponential random graph models. The method also applies to more general Boltzmann distributions, used in statistical physics.

The outline of the paper is as follows. Section \ref{section:overview_bms} introduces the reader to the concept of Bayesian model comparison. A basic description of exponential random graph models is provided in Section \ref{section:ergm}. In Section \ref{section:bayes_pl} we discuss how to perform the adjustments of the pseudolikelihood for ERGMs with the goal to obtain an approximation of the marginal likelihood and in Section \ref{section:BF_approx} we derive an approximation to the Bayes factor. In Section \ref{section:simulation_potts}, we assess the efficiency of the marginal likelihood approximation with a Potts model example for spatial analysis \citep{potts}, where the size of the lattice allows for exact estimation of the marginal likelihood. 
Detailed ERG model selection experiments are presented in Section \ref{section:applications}. We conclude the paper in Section \ref{section:discussion} with final remarks and recommendations to practitioners based on accuracy of evidence and Bayes factor estimates and computational speed. 
The \pkg{Bergm} package for \pkg{R} \citep{caimo2} implements the methodology in this paper. It is available on the CRAN package repository at http://cran.r-project.org/web/packages/Bergm.

%====================================================================================
%====================================================================================
\section{Overview of Bayesian model selection}\label{section:overview_bms}
Consider the countable model set $\models=\{\models_1,\models_2,\models_3,\ldots\}$. The data $y$ are assumed to have been generated by one of the models in that set.
Bayesian model selection aims at calculating the posterior model probability for model $\models_m$, $\pi(\models_m\mid y)$, where it may be of interest to obtain a-posteriori a single most probable model or a subset of likely models.

We associate each model $\models_m$ with a parameter vector $\param_m$. The prior beliefs for each model are expressed through a prior distribution $p(\models_m)$ $(\sum_{m\in\models}p(\models_m)=1)$ and for the parameters within each model through $p(\param_m\mid\models_m)$.
These specifications allow Bayesian inference to proceed by examining the posterior distribution
%%%%%%%%%%%%%%%%%%%%%%%%%%%%%%%%%%%%%%%%%%
\begin{equation*}
\pi(\param_m,\models_m\mid y)\propto f(y\mid\param_m,\models_m) p(\param_m\mid\models_m) p(\models_m).
\end{equation*}
%%%%%%%%%%%%%%%%%%%%%%%%%%%%%%%%%%%%%%%%%%
The within-model posterior appears as $\pi(\param_m\mid y,\models_m)\propto f(y\mid\param_m,\models_m)p(\param_m\mid\models_m)$. The constant of proportionality, termed the \textit{marginal likelihood} or \textit{evidence}, for model $\models_m$ is expressed by
%%%%%%%%%%%%%%%%%%%%%%%%%%%%%%%%%%%%%%%%%%
\begin{equation*}
\pi(y\mid \models_m) = \int_{\paramspa_m} f(y\mid\param_m,\models_m)p(\param_m\mid\models_m)\;\mathrm{d}\param_m,
\end{equation*}
%%%%%%%%%%%%%%%%%%%%%%%%%%%%%%%%%%%%%%%%%%
assuming a prior distribution for $\param_m$ that leads to a marginal likelihood which is finite.
Precise estimation of the above integral is challenging as it involves a high-dimensional integration over a usually complicated and highly variable function, so in most cases the model evidence is not analytically tractable. Knowledge of the evidence is required to deduce the posterior model probability
%%%%%%%%%%%%%%%%%%%%%%%%%%%%%%%%%%%%%%%%%%
\begin{equation*}
\pi(\models_m\mid y) = \frac{\pi(y\mid \models_m)p(\models_m)}{\sum_{j\in M}^{} \pi(y\mid \models_j)p(\models_j)}
\end{equation*}
%%%%%%%%%%%%%%%%%%%%%%%%%%%%%%%%%%%%%%%%%%
using Bayes theorem. The probability $\pi(\models_m\mid y)$ is treated as a measure of uncertainty for model $\models_m$. Comparison of two competing models in the Bayesian setting is performed through the Bayes factor,
%%%%%%%%%%%%%%%%%%%%%%%%%%%%%%%%%%%%%%%%%%
\begin{equation}\label{eqn:BF}
BF_{m,m'}=\frac{\pi(y\mid\models_m)}{\pi(y\mid\models_{m'})}.
\end{equation}
%%%%%%%%%%%%%%%%%%%%%%%%%%%%%%%%%%%%%%%%%%
which provides evidence in favour of model $\models_m$ compared with model $\models_{m'}$. The larger $BF_{m,m'}$ is, the greater the evidence in favor of $\models_{m}$ compared to $\models_{m'}$. A comprehensive review of Bayes factors is presented by \citet{kassr95}.

In this paper we are concerned with approaches based solely on within-model simulation, where the posterior distribution within model $\models_m$ is examined separately for each $m$. Recent reviews comparing popular methods based on MCMC sampling can be found in \cite{wyse_review} as well as in \cite{ardia_review}.

%====================================================================================
%====================================================================================
\section{Exponential random graph models}\label{section:ergm}
Consider the set of all possible graphs on \textit{n} nodes (actors), $\mathcal{Y}$. A $n\times n$ random adjacency matrix \textit{Y} on \textit{n} nodes and a set of edges (relationships) describes the connectivity pattern of a graph that represents the network data. A realisation of $\textit{Y}$ is denoted with $y$ and the presence or absence of an edge (directed or undirected) between the pair of nodes $(i,j)$ is coded as
%%%%%%%%%%%%%%%
\begin{equation*}
  y_{ij}=\begin{cases}
	        1, &\textrm{if }(i,j) \textrm{ are connected,}\\
          0, &\textrm{otherwise.}\\
          \end{cases}
\end{equation*}
%%%%%%%%%%%%%%%
An edge connecting a node to itself is not permitted so $y_{ii} = 0$.

Exponential random graph models represent a general class of models for specifying the probability distribution for a set of random graphs or networks based on exponential-family theory \citep{wasserman}. Local structures in the form of meaningful subgraphs model the global structure of the network. ERGMs model directly the network using the likelihood function
%%%%%%%%%%%%%%%
\begin{equation}\label{eqn:ergmprob}
f(y\mid\param)=\frac{q(y\mid\param)}{z({\param)}}=\frac{\exp\left\{\param^{\top}s(y)\right\}} {\sum_{y\in \mathcal{Y}}^{} \exp\left\{\param^{\top}s(y)\right\}},\qquad \param^{\top}s(y)=\sum_{j=1}^{d}\param_{j}s_{j}(y),
\end{equation}
%%%%%%%%%%%%%%%
where $q(y\mid\param)$ is the  un-normalised likelihood, $s: \mathcal{Y}\rightarrow \mathbb{R}^{d}$ are sufficient statistics based on the adjacency matrix and $\param\in\paramspa\subseteq\mathbb{R}^{d}$ is the vector of model parameters \citep{hunter,snijders_recent_2007}. Our focus lies on ERG models that are edge-dependent, and whose likelihood is intractable.

The evaluation of $z(\param)$ is feasible for only trivially small graphs as this sum involves $2^{\binom{n}{2}}$ terms for undirected graphs. Recent studies on the inference of ERGMs with the Bayesian approach include \cite{koskinen}, \cite{caimo1}, \cite{wang}, \cite{caimo4}, \cite{thiemichen} and \cite{bouranismisp}. Bayesian model selection for exponential random graph models has been explored by \cite{caimo3}, \cite{friel4}, \cite{thiemichen} and \cite{johansen}.

A reparameterization of \eqref{eqn:ergmprob} can express the distribution of the Bernoulli variable $Y_{ij}$ under the conditional form
%%%%%%%%%%%%%%%%%%%%%%%%%%%%%%%%%%
\begin{equation*}
\text{logit}\left\{p(y_{ij}=1\mid y_{-ij},\param)\right\}=\param^{\top}\delta_{s}(y)_{ij},
\end{equation*}
%%%%%%%%%%%%%%%%%%%%%%%%%%%%%%%%%%
where $\delta_{s}(y)_{ij}=s(y^{+}_{ij})-s(y^{-}_{ij})$ denotes the vector of change statistics. The vector is associated with the dyad $y_{ij}$ corresponding to a particular pair of nodes $(i,j)$ and represents the change in the vector of
network sufficient statistics when $y_{ij}$ is toggled from a 0 (no edge, $y^{-}_{ij}$) to a 1 (edge, $y^{+}_{ij}$), holding the rest of the network, $y_{-ij}=y\backslash\{y_{ij}\}$, fixed.
The pseudolikelihood method, developed by \cite{besag2} and applied to social networks by \cite{strauss}, defines an approximation of the full joint distribution in \eqref{eqn:ergmprob} as the product of the full conditionals for individual observations/ dyads:
%%%%%%%%%%%%%%%%%%%%%%%%%%%%%%%%
\begin{equation*}
f^{}_{\text{PL}}(y\mid \param)=\prod_{\substack{i\neq j \\ i< j}}^{}p(y_{ij}\mid y_{-ij},\param)=\prod_{\substack{i\neq j \\ i< j}}^{}\frac{p(y_{ij}=1\mid y_{-ij},\param)^{y_{ij}}}{\{1-p(y_{ij}=1\mid y_{-ij},\param)\}^{y_{ij}-1}},
\end{equation*}
%%%%%%%%%%%%%%%%%%%%%%%%%%%%%%%%
where $\text{y}_{-ij}$ denotes $\text{y}\backslash\{\text{y}_{ij}\}$. The condition $i< j$ holds for undirected graphs.

%====================================================================================
%====================================================================================
\section{Adjusting the pseudolikelihood}\label{section:bayes_pl}
Analytical or computational intractability of the likelihood function poses a major challenge to Bayesian inference, as well as to all likelihood-based inferential approaches. A natural strategy to deal with such model intractability is to substitute the full likelihood with a surrogate composite likelihood \citep{lindsay,varin1}, that shares similar properties with the full likelihood. The pseudolikelihood \citep{besag1,besag2} is a special case of the composite likelihood and can serve as a proxy to the full likelihood when the assumption of conditional independence of the variables is reasonable.

This assumption is usually unrealistic, though. The drawback of the pseudolikelihood is that it ignores strong dependencies in the data and can, therefore, lead to a biased estimation. We propose to perform adjustments on the pseudolikelihood to obtain a reasonable approximation to the likelihood.

Adjustments to composite likelihood functions have been previously suggested by \cite{ribatet}. Following the proposals of \cite{friel1} and \cite{bouranismisp} for GRFs, we initially adjust the pseudolikelihood itself by matching its first two moments with the first two moments of the likelihood through a model-specific invertible and differentiable mapping
%%%%%%%%%%%%%%%%%%%%%%%%%%%%%%%%%%%%%%%%%%
\begin{equation}\label{eq:mode_curv}
  g:\begin{cases}
       \paramspa\rightarrow\paramspa\\
       \param \mapsto \htheta_{MPLE}+W(\param-\htheta_{MLE}),
      \end{cases}
\end{equation}
%%%%%%%%%%%%%%%%%%%%%%%%%%%%%%%%%%%%%%%%%%
which depends on the maximum likelihood estimate, $\htheta_{MLE}$, the maximum pseudolikelihood estimate, $\htheta_{MPLE}$, and a transformation matrix $W$. The \textit{mode and curvature-adjusted pseudolikelihood} is defined as the function $y\mapsto f^{}_{\text{PL}}(y\mid g(\param))$.

Figure \ref{fig:correction_graph} displays a difference in magnitude from $f(y\mid \param)$; a magnitude adjustment of the mode and curvature-adjusted pseudolikelihood results in the \textit{fully adjusted pseudolikelihood}
%%%%%%%%%%%%%%%%%%%%%%%%%%%%%%%%%%%%%%%%%%
\begin{equation}\label{eq:fully_adj}
\tilde{f}(y\mid \param)=C\cdot f^{}_{\text{PL}}(y\mid g(\param))\,,
\end{equation}
%%%%%%%%%%%%%%%%%%%%%%%%%%%%%%%%%%%%%%%%%%
for some constant $C>0$. The remainder of this section provides guidelines for estimating $C$ and obtaining the mapping $g$ through the estimations of $\htheta_{MPLE}$, $\htheta_{MLE}$, and $W$.

\subsection{Mode adjustment}
Empirical analysis by \cite{friel1} and \cite{bouranismisp} showed that the Bayesian estimators resulting from using the pseudolikelihood function as a plug-in for the true likelihood function are biased and their variance can be underestimated. It is, therefore, natural to consider a correction of the mode of the pseudolikelihood approximation.

Paramount to the approach is the ability to estimate the maxima of the likelihood and the pseudolikelihood,
%%%%%%%%%%%%%%%%%%%%%%%%%%%%%%%%%%%%%%%%%%
\begin{align}\label{eq:mode_est}
\htheta_{MLE}  &= \argmax_{\theta}\log{f(y\mid\param)},\\
\htheta_{MPLE} &= \argmax_{\theta}\log{f^{}_{\text{PL}}(y\mid\param)}.\nonumber
\end{align}
%%%%%%%%%%%%%%%%%%%%%%%%%%%%%%%%%%%%%%%%%%
While the MPLE is fast and straightforward to obtain because of the closed form of the pseudolikelihood, care is needed when estimating the MLE for ERGMs. We considered the Monte Carlo Maximum Likelihood Estimation (MC-MLE) procedure proposed by \cite{geyer}. Alternative procedures exist, see \cite{hunter}.

%MCMC estimation procedures \citep{Snijders02} can be considered for approximating a solution to \eqref{eq:mode_est}

\subsection{Curvature adjustment}
Composite likelihoods have been previously shown to modify the correlation between the variables \citep{friel1}.
The mapping in \eqref{eq:mode_curv} ensures that the adjusted pseudolikelihood and the full likelihood have the same mode and aims to recover the overall geometry of the distribution (Figure \ref{fig:correction_graph}). We choose the transformation matrix $W$ that satisfies
%%%%%%%%%%%%%%%%%%%%%%%%%%%%%%%%%%%%%%%%%%
\begin{align}\label{eq:mode_adj}
%\nabla_{\param}\log{f(y\mid\param)}|_{\htheta_{MLE}}& = W^{\top}\nabla_{\param}\log{f^{}_{\text{PL}}(y\mid\param)}|_{\htheta_{MPLE}}\nonumber\\
\nabla^{2}_{\param}\log{f(y\mid\param)}|_{\htheta_{MLE}} = W^{\top}\nabla^{2}_{\param}\log{f^{}_{\text{PL}}(y\mid\param)}|_{\htheta_{MPLE}}W,
\end{align}
%%%%%%%%%%%%%%%%%%%%%%%%%%%%%%%%%%%%%%%%%%
so that the gradient and the Hessian of the log-likelihood and $\log\tilde{f}(y\mid \param)$ are the same. It is possible to estimate the gradient and the Hessian of the log-likelihood using the following two identities:
%%%%%%%%%%%%%%%%%%%%%%%%%%%%%%%%%%%%%%%%%%
\begin{align*}
\nabla_{\param}\log{f(y\mid\param)}
&=s(y)-\frac{z'(\param)}{z(\param)}\nonumber\\
&=s(y)-\frac{\sum s(y)\text{exp}\left\{\param^{\top} s(y)\right\}}{\sum\text{exp}\left\{\param^{\top} s(y)\right\}}\nonumber\\
&=s(y)-\mathbb{E}_{y\mid \param}\left[s(y)\right],
\end{align*}
%%%%%%%%%%%%%%%%%%%%%%%%%%%%%%%%%%%%%%%%%%
and
%%%%%%%%%%%%%%%%%%%%%%%%%%%%%%%%%%%%%%%%%%
\begin{align*}
\nabla^{2}_{\param}\log{f(y\mid\param)}
&=\nabla_{\param}\left[-\frac{z'(\param)}{z(\param)}\right]=-\frac{z^{\prime\prime}(\param)z(\param)-z'(\param)z'(\param)}{z^2(\param)}\nonumber\\
&=-\bigg\{\mathbb{E}_{y\mid \param}\left[s^2(y)\right]-\left[\mathbb{E}_{y\mid \param}\left[s(y)\right]\right]^2\bigg\}\nonumber\\
&=-\mathbb{V}_{y\mid \param}\left[s(y)\right],
\end{align*}
%%%%%%%%%%%%%%%%%%%%%%%%%%%%%%%%%%%%%%%%%%
where $\mathbb{V}_{y\mid \param}\left[s(y)\right]$ denotes the covariance matrix of $s(y)$ with respect to $f(y\mid\param)$. The presence of the normalising term renders exact evaluation of $\mathbb{E}_{y\mid \param}\left[s(y)\right]$ and $\mathbb{V}_{y\mid \param}\left[s(y)\right]$ intractable. We resort to Monte Carlo sampling from $f(y\mid\param)$ in order to estimate these.

The Hessian matrices at the maximum of their respective function are negative definite matrices and therefore admit a Cholesky decomposition,
%%%%%%%%%%%%%%%%%%%%%%%%%%%%%%%%%%%%%%%%%%
\begin{align}\label{eq:curv_adj}
-\nabla^{2}_{\param}\log{f(y\mid\param)}|_{\htheta_{MLE}} &= N^{\top} N \nonumber\\
-\nabla^{2}_{\param}\log{f^{}_{\text{PL}}(y\mid\param)}|_{\htheta_{MPLE}}&=M^{\top} M,
\end{align}
%%%%%%%%%%%%%%%%%%%%%%%%%%%%%%%%%%%%%%%%%%
where $M$ and $N$ are upper triangular matrices with strictly positive diagonal entries. By straightforward algebra, combining \eqref{eq:mode_adj} and \eqref{eq:curv_adj} yields $W=M^{-1} N$.

\subsection{Magnitude adjustment}
The magnitude adjustment aims to scale the mode and curvature-adjusted pseudolikelihood to the appropriate magnitude by performing a linear transformation of the vertical axis. The constant $C$ in \eqref{eq:fully_adj} is defined so that $\tilde{f}(y\mid \htheta_{MLE}) = f(y\mid \htheta_{MLE})$, which implies
%%%%%%%%%%%%%%%%%%%%%%%%%%%%%%%%%%%%%%%%%%
\begin{equation}
C=\frac{q(y\mid\htheta_{MLE})\cdot z^{-1}(\htheta_{MLE})}{f^{}_{\text{PL}}(y\mid g(\htheta_{MLE}))}.
\label{eq:mult_const}
\end{equation}
%%%%%%%%%%%%%%%%%%%%%%%%%%%%%%%%%%%%%%%%%%
Since $z(\htheta_{MLE})$ is intractable, we unbiasedly estimate $C$ by replacing the normalising constant with an estimator which we now describe, following \cite{friel4}. We introduce an auxiliary variable $t\in[0,1]$ discretised as $0=t^{}_0<t^{}_1<\ldots<t^{}_L=1$ and consider the distributions
%%%%%%%%%%%%%%%%%%%%%%%%%%%%%%%%%%%%%%%%%%%
\begin{equation*}
f(y\mid t_j\param)
   = \frac{q(y\mid t_j\param)}
          {z(t_j\param)}
   = \frac{\exp\{(t_j\param)^{\top} s(y)\}}
          {\sum_{y \in \mathcal{Y}} \exp\{(t_j\param)^{\top} s(y)\}},~~j\in\{0,\ldots,L\}.
\end{equation*}
%%%%%%%%%%%%%%%%%%%%%%%%%%%%%%%%%%%%%%%%%%%
An estimate of $z(\htheta_{MLE})$ can be obtained using
%%%%%%%%%%%%%%%%%%%%%%%%%%%%%%%%%%%%%%%%%%
\begin{equation}\label{eq:exp_path}
\frac{z(\htheta_{MLE})}{z(0)}=\frac{z(t_L\htheta_{MLE})}{z(t_0\htheta_{MLE})}=\prod_{j=0}^{L-1}\frac{ z(t_{j+1}\htheta_{MLE}) }{ z(t_j\htheta_{MLE}) },
\end{equation}
%%%%%%%%%%%%%%%%%%%%%%%%%%%%%%%%%%%%%%%%%%
where $z(0)=2^{\binom{n}{2}}$ for undirected graphs and $n$ is the number of nodes. Note that in the case of a Potts/autologistic model, $z(0)=2^N$, where $N$ is the size of the lattice.

Importance sampling is used to estimate the ratios of normalising constants in \eqref{eq:exp_path}. We take the un-normalised likelihood $q(y\mid t_j\param)$ as an importance distribution for the "target" distribution $f(y\mid t_j\param)$, noting that
%%%%%%%%%%%%%%%%%%%%%%%%%%%%%%%%%%%%%%%%%%%
\begin{equation*}
\frac{ z(t_{j+1}\htheta_{MLE}) }{ z(t_j\htheta_{MLE}) }=\mathbb{E}^{}_{y\mid t_j\htheta_{MLE}}\left[\frac{ q(y\mid t_{j+1}\htheta_{MLE}) }{ q(y\mid t_j\htheta_{MLE})) }\right].
\end{equation*}
%%%%%%%%%%%%%%%%%%%%%%%%%%%%%%%%%%%%%%%%%%%
An unbiased importance sampling estimate of this expectation can be obtained by simulating multiple draws $y^{(j)}_1,\dots,y^{(j)}_K \sim f(y\mid t_j\htheta_{MLE})$, yielding
%%%%%%%%%%%%%%%%%%%%%%%%%%%%%%%%%%%%%%%%%%
\begin{equation*}
\widehat{\frac{ z(t_{j+1}\htheta_{MLE}) }{ z(t_j\htheta_{MLE}) }}=\frac{1}{K}\sum_{k=1}^{K}\frac{q(y^{(j)}_k\mid t_{j+1}\htheta_{MLE})}{q(y^{(j)}_k\mid t_j\htheta_{MLE})}.
\end{equation*}
%%%%%%%%%%%%%%%%%%%%%%%%%%%%%%%%%%%%%%%%%%
Increasing the number of temperatures $L$ and the number of simulated graphs will lead to a more precise estimate of $z(\htheta_{MLE})$ and will necessarily increase the computational burden. However, as we 
shortly illustrate, however, this does not add significantly to the overall computational cost of the adjustment procedure. We note that estimation of $z(\htheta_{MLE})$ is performed once upfront for each competing model. The estimate of $C$ is
%%%%%%%%%%%%%%%%%%%%%%%%%%%%%%%%%%%%%%%%%%
\begin{equation*}
\hat{C}=\frac{q(y\mid\htheta_{MLE})\cdot \hat{z}^{-1}(\htheta_{MLE})}{f^{}_{\text{PL}}(y\mid g(\htheta_{MLE}))},
\end{equation*}
%%%%%%%%%%%%%%%%%%%%%%%%%%%%%%%%%%%%%%%%%%
where $\hat{z}(\htheta_{MLE})$ follows from \eqref{eq:exp_path}.

We note that estimators similar to \eqref{eq:exp_path} could be obtained using annealed importance sampling \citep{neal} or a sequential Monte Carlo algorithm \citep{delmoral}. Indeed, these approaches do not require independent simulations from the likelihood for different temperatures and may therefore provide greater accuracy per computational cost and could be considered as alternatives. However, depending on the implementation of those methods, the resulting estimator may be biased.

%====================================================================================
%====================================================================================
\section{Approximation of the Bayes factor}\label{section:BF_approx}
Replacing the likelihood with the unadjusted pseudolikelihood approximation in Bayes formula yields the within-model pseudo-posterior distribution
%%%%%%%%%%%%%%%%%%%%%%%%%%%%%%%%%%%%%%%%%%
\begin{equation}\label{eqn:pseudopostdistr}
\pi^{}_{\text{PL}}(\param_m\mid y,\models_m)=\frac{f^{}_{\text{PL}}(y\mid\param_m,\models_m) p(\param_m\mid\models_m)}{\pi^{}_{\text{PL}}(y\mid \models_m)}=\frac{f^{}_{\text{PL}}(y\mid\param_m,\models_m) p(\param_m\mid\models_m)}{\int_{\paramspa_m} f^{}_{\text{PL}}(y\mid\param_m,\models_m)p(\param_m\mid\models_m)\;\mathrm{d}\param_m}.
\end{equation}
%%%%%%%%%%%%%%%%%%%%%%%%%%%%%%%%%%%%%%%%%%
In analogy to \eqref{eqn:BF}, the Bayes factor based on the unadjusted pseudolikelihood approximation is
%%%%%%%%%%%%%%%%%%%%%%%%%%%%%%%%%%%%%%%%%%
\begin{equation*}
BF^{PL}_{mm'}= \frac{\pi^{}_{\text{PL}}(y\mid \models_m)}{\pi^{}_{\text{PL}}(y\mid \models_{m'})}=
\frac{
\int_{\paramspa_m}^{}f^{}_{\text{PL}}(y\mid\param_m,\models_m)p(\param_m\mid\models_m)\;\mathrm{d}\param_m}{
\int_{\paramspa_{m'}}^{}f^{}_{\text{PL}}(y\mid\param_{m'},\models_{m'})p(\param_{m'}\mid\models_{m'})\;\mathrm{d}\param_{m'}}.
\end{equation*}
%%%%%%%%%%%%%%%%%%%%%%%%%%%%%%%%%%%%%%%%%%
A naive implementation of the pseudolikelihood or any other higher-order composite likelihood is likely to give misleading marginal likelihood estimates, as we illustrate in Section \ref{section:simulation_potts}. Having completed the adjustment steps, we propose to approximate the within-model posterior distribution by
%%%%%%%%%%%%%%%%%%%%%%%%%%%%%%%%%%%%%%%%%%
\begin{equation}\label{eq:cor_pp}
\widetilde{\pi}(\param\mid y,\models_m)=
\frac{\tilde{f}(y\mid \param_m,\models_m)p(\param_m\mid\models_m)}{\widetilde{\pi}(y\mid \models_m)}=
\frac{\tilde{f}(y\mid \param_m,\models_m)p(\param_m\mid\models_m)}{\int_{\paramspa_m} \tilde{f}(y\mid \param_m,\models_m)p(\param_m\mid\models_m)\;\mathrm{d}\param_m},
\end{equation}
%%%%%%%%%%%%%%%%%%%%%%%%%%%%%%%%%%%%%%%%%%
Working with \eqref{eq:cor_pp} we can now approximate \eqref{eqn:BF} by
%%%%%%%%%%%%%%%%%%%%%%%%%%%%%%%%%%%%%%%%%%
\begin{equation*}
\widetilde{BF}_{mm'} = \frac{\widetilde{\pi}(y\mid \models_m)}{\widetilde{\pi}(y\mid \models_{m'})}=
\frac{\int_{\paramspa_m} \tilde{f}(y\mid \param_m,\models_m)p(\param_m\mid\models_m)\;\mathrm{d}\param_m}{\int_{\paramspa_{m'}} \tilde{f}(y\mid \param_{m'},\models_{m'})p(\param_{m'}\mid\models_{m'})\;\mathrm{d}\param_{m'}}.
\end{equation*}
%%%%%%%%%%%%%%%%%%%%%%%%%%%%%%%%%%%%%%%%%%
The aforementioned framework offers one possibility in the Bayesian setting to obtain an approximation to the within-model posterior distribution. Another possibility has been explored by \cite{bouranismisp}, whose ERGM experiments showed that estimation with the pseudo-posterior distribution is biased. The authors presented an algorithm to draw an approximate sample from the intractable posterior distribution $\pi(\param\mid y)$.

Their suggested approach first samples from the pseudo-posterior distribution \eqref{eqn:pseudopostdistr}.
Then an invertible and differentiable mapping $\phi:\paramspa\rightarrow\paramspa$ is considered to transform the entire sample $\{\param_i\}^{T}_{i=1}$ so that it is a sample from an approximation of the posterior distribution, whose density is
%%%%%%%%%%%%%%%%%%%%%%%%%%%%%%%%%%%%%%%%%%
\begin{equation*}
\hat{\pi}(\param\mid y, \models_m)=\pi^{}_{\text{PL}}(\phi^{-1}(\param)\mid y,\models_m)\cdot \left|\frac{\partial \phi^{-1}(\param)}{\partial \param}\right|.
\end{equation*}
%%%%%%%%%%%%%%%%%%%%%%%%%%%%%%%%%%%%%%%%%%
Following this approach and applying a change of variables, the model evidence $\pi(y\mid \models_m)$ is approximated by
%%%%%%%%%%%%%%%%%%%%%%%%%%%%%%%%%%%%%%%%%%
\begin{equation*}
\hat{\pi}(y\mid\models_m)=\int_{\paramspa}\pi^{}_{\text{PL}}(\phi^{-1}(\param),y\mid \models_m)\left|\frac{\partial \phi^{-1}(\param)}{\partial \param}\right|\;\mathrm{d}\param = \pi^{}_{\text{PL}}(y\mid \models_m)
\end{equation*}
%%%%%%%%%%%%%%%%%%%%%%%%%%%%%%%%%%%%%%%%%%
and it follows that there is no gain from the transformation of the pseudo-posterior distribution when the aim is to obtain a reasonable approximation of the marginal likelihood. As such, while the correction algorithm of \cite{bouranismisp} is appropriate for conducting Bayesian inference on the model parameters, it is not suitable for model selection.

% {\color{red}Finally, we explain why the factor $C$ is needed to estimate the evidence. To see this, let $\pi^{}_{\text{MC}}(\param\mid y,\models_m)$ be the posterior distribution resulting from the mode and curvature-adjusted pseudolikelihood, $f^{}_{\text{PL}}(y\mid g(\param),\models_m)$. Equation \eqref{eq:fully_adj} shows that the relationship between $\widetilde{\pi}(\param\mid y,\models_m)$ and $\pi^{}_{\text{MC}}(\param\mid y,\models_m)$ is
% %%%%%%%%%%%%%%%%%%%%%%%%%%%%%%%%%%%%%%%%%%
% \begin{align*}
% \widetilde{\pi}(\param\mid y,\models_m)&=
% \frac{C\cdot f^{}_{\text{PL}}(y\mid g(\param_m),\models_m)p(\param_m\mid\models_m)}{\int_{\paramspa_m} C\cdot f^{}_{\text{PL}}(y\mid g(\param_m),\models_m)p(\param_m\mid\models_m)\;\mathrm{d}\param_m}\\
% &=
% \frac{f^{}_{\text{PL}}(y\mid g(\param_m),\models_m)p(\param_m\mid\models_m)}{\int_{\paramspa_m} f^{}_{\text{PL}}(y\mid g(\param_m),\models_m)p(\param_m\mid\models_m)\;\mathrm{d}\param_m}\\
% &=
% \pi^{}_{\text{MC}}(\param\mid y,\models_m).
% \end{align*}
% %%%%%%%%%%%%%%%%%%%%%%%%%%%%%%%%%%%%%%%%%%
% Therefore, the two posteriors coincide, but their marginal likelihoods differ.}

%====================================================================================
%====================================================================================
\section{Potts simulation study}\label{section:simulation_potts}
% Use the slides in http://informs-sim.org/wsc12papers/includes/files/inv172.pdf
The Ising model has been a popular approach to modeling spatial binary data $y = \{y_1,\ldots,y_N\}\in\{-1;1\}^{N}$ on a lattice of size $N=\upsilon\times \nu$, where $\upsilon$ and $\nu$ are the number of rows (height) and columns (width) of the lattice, respectively. 
A lattice with $N$ nodes has $2^N$ possible realizations; the normalising constant $z(\param)$ in \eqref {intro_eqn:likelihood} is a summation over all of the realizations and it becomes analytically unknown for moderate sized graphs. The autologistic model \citep{besag3} extends the Ising model to allow for unequal abundances of each state value, while the Potts model \citep{potts} allows each lattice point to take one of $S\geq 2$ possible values/states.

\begin{figure}[H]
\centering
\begin{tikzpicture}[scale=0.9]
\foreach \x in {0,...,4}{
 \draw[line width=1.1pt, black] (\x,0) to[out=90,in=-90] (\x,4);
  \draw[line width=1.1pt, black] (0,\x) to[out=0,in=180] (4,\x);
}
\foreach \x in {0,...,4}
	\foreach \y in {0,...,4}
   		\draw[fill = gray!5] (\x,\y) circle (1.7mm);

\draw[fill = red]  (2,2) circle (1.7mm);
\draw[fill = blue] (2,3) circle (1.7mm);
\draw[fill = blue] (1,2) circle (1.7mm);
\draw[fill = blue] (3,2) circle (1.7mm);
\draw[fill = blue] (2,1) circle (1.7mm);
\end{tikzpicture}
\caption{Example of a first-order neighborhood graph. The closest neighbors of the node in red are represented by nodes in blue.}
\label{fig:ising_neih}
\end{figure}
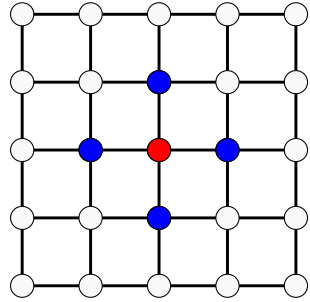

In this example we investigate the efficiency of the approximation to the marginal likelihood when the likelihood is replaced by $\tilde{f}(y\mid \param_m,\models_m)$, with a small dataset for which we can carry out exact computations. 30 realizations from an isotropic 2-state Potts model with interaction parameter $\param=0.4$ defined on a lattice of size $15\times 15$ were exactly sampled via \cite{rue} and \cite{GiRaF}. The sufficient statistic for the Potts model is the number of corresponding neighbors in the graph
%%%%%%%%%%%%%%%%%%%%%%%%%%%%%%%%%%%%%%%%%%
\[
s(y) = \sum_{j<i}\sum_{j\sim i} \mathbbm{1}\{y_i=y_j\},
\]
%%%%%%%%%%%%%%%%%%%%%%%%%%%%%%%%%%%%%%%%%%
where the notation $j\sim i$ denotes that node $j$ is a neighbor of node $i$. We assume that the lattice points have been indexed from top to bottom in each column and that columns are ordered from left to right.
For a first-order neighborhood model an interior point $y_i$ has neighbors $\{y_{i-\upsilon}, y_{i-1}, y_{i+1},y_{i+\upsilon}\}$; nodes situated at the boundary of the grid have less than four neighbors (Figure \ref{fig:ising_neih}). 
%%%%%%%%%%%%%%%%%%%%%%%%%%%%%%%%%%%%%%%%%%
%%%%%%%%%%%%%%%%%%%%%%%%%%%%%%%%%%%%%%%%%%
The model under consideration is expressed as
%%%%%%%%%%%%%%%%%%%%%%%%%%%%%%%%%%%%%%%%%%
\begin{equation}\label{eqn:ising_post}
\pi(\param\mid y)\propto z(\param)^{-1}\exp\left\{\param^{\top} s(y)\right\}p(\param),
\end{equation}
%%%%%%%%%%%%%%%%%%%%%%%%%%%%%%%%%%%%%%%%%%
where a diffuse Gaussian prior distribution, $\mathcal{N}\left(0,25\right)$, was assumed.

The size of the simulated lattices allows for accurate estimates of the evidence as follows: the normalising constant $z(\param)$ can be calculated exactly with a recursive forward-backward algorithm \citep{reeves, rue}, which can then be plugged into the right hand side of \eqref{eqn:ising_post}. Numerical integration (using the trapezoidal rule) of the right hand side over a grid of $\{\param_i\}_{i=1}^{M}$ values gives an accurate estimate of $\pi(y)$:
%%%%%%%%%%%%%%%%%%%%%%%%%%%%%%%%%%%%%%%%%%
\begin{equation*}
\hat{\pi}(y)= \sum_{i=2}^M \dfrac{(\param_i-\param_{i-1})}{2}
\left[\dfrac{q(y\mid\param_i)p(\param_i)}{z(\param_i)}+\dfrac{q(y\mid\param_{i-1})p(\param_{i-1})}{z(\param_{i-1})}\right].
\end{equation*}
%%%%%%%%%%%%%%%%%%%%%%%%%%%%%%%%%%%%%%%%%%
This serves as a ground truth against which to compare with the corresponding estimates of the model evidence under the fully adjusted pseudolikelihood, detailed in Section \ref{section:bayes_pl}. To reduce discretisation error we considered a sequence of length 5,000 over the interval [0, 0.8], covering the effective range of values that $\param$ can take. The same numerical integration can be used to estimate the evidence based on the (corrected) pseudo-posterior distribution.

In the case of the Potts model the pseudolikelihood comprises the product of full-conditional distributions of each $y_{i}$,
%%%%%%%%%%%%%%%%%%%%%%%%%%%%%%%%%%%%%%%%%%
\begin{equation*}
f^{}_{\text{PL}}(y\mid \param)=\prod_{i=1}^{N}f(y_{i}\mid y_{-i},\param),
\end{equation*}
%%%%%%%%%%%%%%%%%%%%%%%%%%%%%%%%%%%%%%%%%%
where $y_{-i}$ denotes $y\backslash\{\text{y}_{i}\}$. For a lattice of size $N$ the denominator of the pseudo-likelihood will take $2N\ll 2^{N}$ calculations to evaluate, providing a significant improvement when compared to the full likelihood. When $\param=0$, the pseudolikelihood function is identical to the true likelihood.

In this example we estimated \eqref{eq:exp_path} using a ladder of 100 equally spaced path points and sampling 1500 graph statistics at each of them. The algorithm took 1.5 min to estimate the intractable normalising constant for each dataset. This setup has been empirically shown to be sufficiently accurate in this example, where the estimate of the intractable normalising constant at the mode agrees with the respective estimate using the recursive forward-backward algorithm.
%%%%%%%%%%%%%%%%%%%%%%%%%%%%%%%%%%%%%%%%%%
\begin{figure}[H]
\centering
\vspace{-5.6em}
\includegraphics[width=16cm,height=13cm]{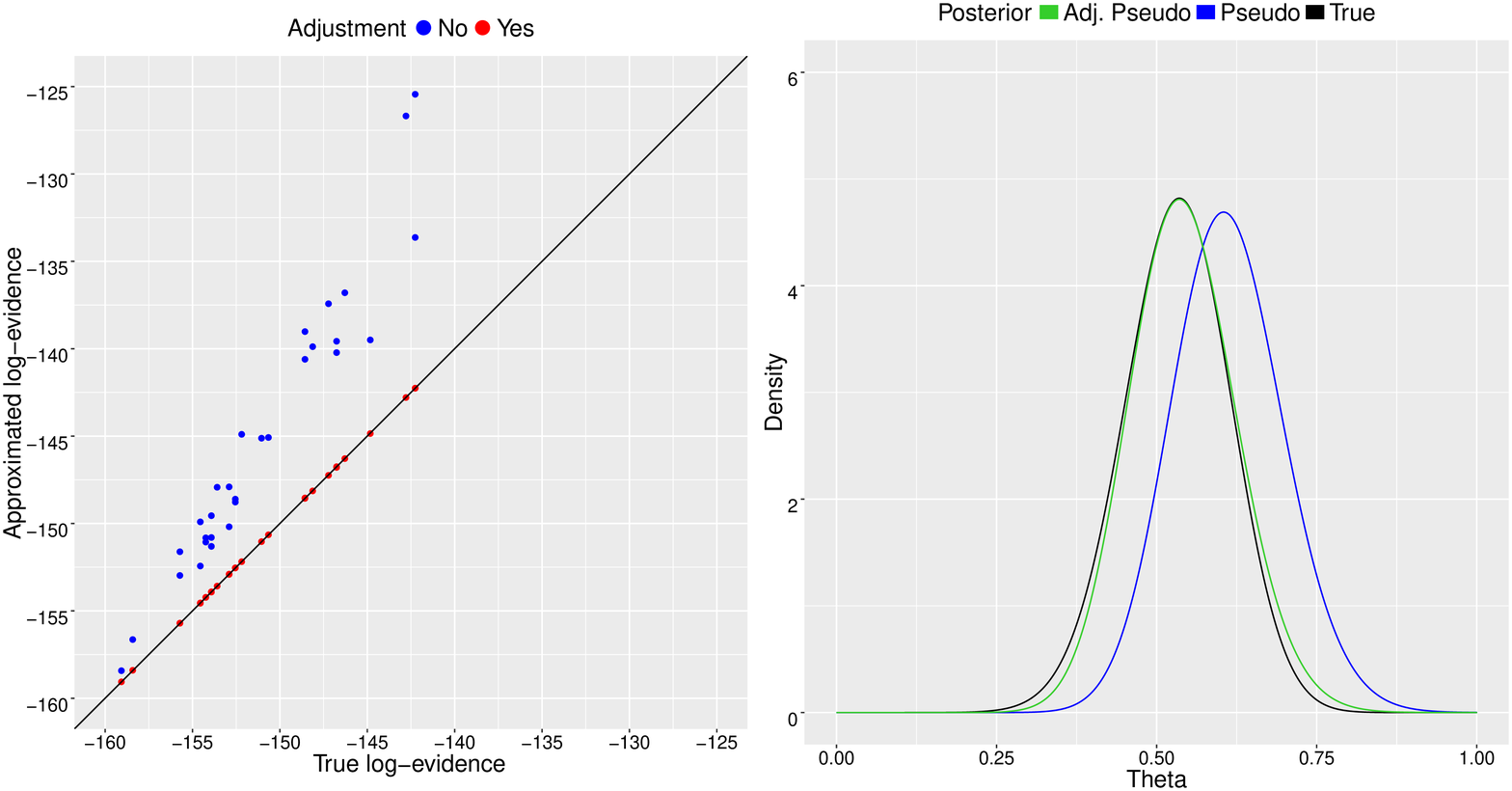}
\vspace{-5.5em}
\caption{Estimates of the approximate log-evidence based on the adjusted pseudolikelihood and the unadjusted pseudolikelihood against the true value of the log-evidence for 30 datasets (left panel).
Blue points correspond to $\log\pi^{}_{\text{PL}}(y)$ and red points correspond to $\log\widetilde{\pi}(y)$. Right panel: normalised posterior density for a dataset with strong dependence structure and log-evidence estimates $(\log\pi^{}_{\text{PL}}(y),\log\widetilde{\pi}(y))=(-125.46,-142.25)$.
This shows that $\widetilde{\pi}(\param\mid y)$, based on the adjusted pseudolikelihood, is very similar to the true posterior distribution.}
\label{fig:potts_model_graphs}
\end{figure}
%%%%%%%%%%%%%%%%%%%%%%%%%%%%%%%%%%%%%%%%%%
In order to assess the accuracy of the approximation to the marginal likelihood based on the fully adjusted pseudolikelihood, in Figure \ref{fig:potts_model_graphs} we present a scatterplot of the
true value of $\log \pi(y)$ against $\log\pi^{}_{\text{PL}}(y)$ and $\log\widetilde{\pi}(y)$ for a range of scenarios with different dependence structures. For datasets with strong dependence structure, the unadjusted pseudolikelihood is inefficient and it should be avoided for model comparisons.

On the contrary, an approximation based on the fully adjusted pseudolikelihood gives excellent performance irrespectively of the dependence structure and is virtually an exact estimate to the true evidence. We note that when we are able  to calculate $z(\htheta_{MLE})$ exactly using the recursive forward-backward algorithm we have almost exact agreement between $\log \pi(y)$ and $\log\widetilde{\pi}(y)$.

%====================================================================================
%====================================================================================
\section{Applications}\label{section:applications}
The intractability of the ERG likelihood function leads to a doubly-intractable posterior distribution. Due to the intractable normalizing constant of the likelihood function, the estimation of the evidence is added with another layer of difficulty. 
Our tractable approximation to the likelihood yields an approximated posterior distribution, making it possible to use existing evidence estimation techniques in the presence of intractable multi-dimensional integrals (see Section \ref{section:intro}).

Such within-model strategies focus on the (approximated) posterior distribution for each competing model $\models_m$ separately, with the aim to estimate their model evidence. For illustration purposes, we considered three kinds of strategies that are based on MCMC simulation: Chib and Jeliazkov's one block Metropolis-Hastings method, thermodynamic integration (TI) and Steppingstone sampling.
Below is a brief overview of these and a more detailed presentation is available at the Supplementary material. Algorithm \ref{alg:model_evidence} summarises the steps involved in estimating the evidence $\widetilde{\pi}(y\mid\models_m)$ for each model.
%%%%%%%%%%%%%%%%%%%%%%%%%%%%%%%%%%%%%%%%%%
\begin{itemize}
	\item \textbf{One block Metropolis:} equation \eqref{eq:posterior} is rearranged with respect to the marginal likelihood.
	Assuming a tractable likelihood, estimation of the marginal likelihood requires an estimate of the posterior ordinate $\pi(\param\mid y)$, which can be found at a high-density point $\param^{*}$ in the support of the target posterior distribution for estimation efficiency.
The method is used in the case where the parameter vector $\param$ can be updated in a single block and has been generalised to situations where the parameter vector is updated in full-conditional blocks \citep{chib2}.
	\item \textbf{Power posteriors and controlled thermodynamic integration (CTI):} a path sampling type method and an application of the thermodynamic integration (TI) technique from statistical physics \citep{friel5}. The power posterior distribution is defined to be proportional to the product of the prior and the likelihood raised to a power $t$ where $t\in [0,1]$ so that $\pi_{t}(\param\mid y) \propto f(y\mid \param)^t p(\param)$. The inverse temperature $t\in [0,1]$ has the effect of tempering the likelihood. A sample is drawn from the power posterior distribution and it is used to estimate the intractable evidence. An improvement on the variance of the evidence estimator can be achieved through the use of control variates \citep{oates}.
	\item \textbf{Stepping stones sampler:} uses the idea of powered posteriors, treating them as a series of intermediate distributions between the prior and the posterior. The corresponding normalising constant of the power posterior is $z(y\mid t) = \int_{\param}f(y\mid \param)^t p(\param)\;\mathrm{d}\param$. An estimate of the evidence, $z(y\mid t_m=1)$ is given by the product of $k=0,\ldots,m-1$ ratios of consecutive normalising constants, $r_{k}=z(y\mid t_{k+1})/z(y\mid t_{k})$, see \cite{xie}.
\end{itemize}
%%%%%%%%%%%%%%%%%%%%%%%%%%%%%%%%%%%%%%%%%%
\begin{algorithm}[H]
\caption{Within-model search framework for evidence estimation}
\label{alg:model_evidence}
\begin{spacing}{1.1}
\begin{algorithmic}[1]
\For{$m~\text{in}~\models=\{\models_1,\models_2,\models_3,\ldots\}$}
	\vspace{.10cm}
	\item[]\hspace{1.55em}\textbf{Adjustment phase to obtain} $\pmb{\widetilde\pi}$
  \vspace{.10cm}
	\State Estimate $\htheta_{MLE,m}$ and $\htheta_{MPLE,m}$ using \eqref{eq:mode_est}.
	\State Estimate $W_m$ based on \eqref{eq:mode_adj}.
	\State Construct $g_m(\param_m)=\htheta_{MPLE,m}+W_m(\param_m-\htheta_{MLE,m})$ to obtain $f^{}_{\text{PL}}(y\mid g_m(\param_m),\models_m)$.
	\State \parbox[t]{\dimexpr\linewidth-\algorithmicindent}{Perform a magnitude adjustment based on \eqref{eq:mult_const} to estimate $C$ and, therefore, obtain \\ $\tilde{f}(y\mid \param_m,\models_m)=C\cdot f^{}_{\text{PL}}(y\mid g_m(\param_m),\models_m)$.\strut}
	%\State MCMC sampling from $\widetilde\pi(\param_m\mid y,\models_m)\propto\tilde{f}(y\mid \param_m,\models_m)p(\param_m\mid\models_m)$.
	\vspace{.10cm}
  \item[]\hspace{1.55em}\textbf{Evidence estimation for} $\pmb{\widetilde\pi}$
  \vspace{.10cm}
  \State Employ a within-model strategy to estimate $\widetilde{\pi}(y\mid \models_m)$.
\EndFor
\end{algorithmic}
\end{spacing}
\end{algorithm}
%%%%%%%%%%%%%%%%%%%%%%%%%%%%%%%%%%%%%%%%%%
Computation in this paper was carried out with the statistical environment R \citep{rsoft} on a laptop computer with an Intel \textregistered Core\textsuperscript{TM} i7-4500U CPU (1.80GHz) and 16GB RAM. Throughout the analysis of the network data we assumed a diffuse Multivariate Gaussian prior distribution for the model parameters, $\mathcal{MVN}\left(0_{d},100\times I_d\right)$, where $0_{d}$ is the null vector and $I_{d}$ is the identity matrix of size equal to the number of model dimensions, $d$, unless stated otherwise.

\subsection{Benchmark algorithms}
Comparisons are provided against the \textit{auto-reversible jump (Auto-RJ) exchange} \citep{caimo3} and the \textit{population exchange} \citep{friel4}.

\begin{itemize}
\item The Auto-RJ is a trans-dimensional MCMC algorithm. The algorithm consists of two steps: the first (offline) step is used to sample from the posterior distribution of each competing model using the exchange algorithm \citep{caimo1} and then to approximate the estimated posterior by
Gaussian distributions determined by the first moments of each sample. 
The second (online) step of the algorithm makes use of the Gaussian posterior proposal estimated in the offline step as within-model proposals for the reversible jump-MCMC computation. 
Note that the Auto-RJ exchange requires draws from the likelihood. Since it is not possible to draw exactly from it, an auxiliary Markov chain from the tie-no-tie (TNT) sampler \citep{ergm} is used to return a draw that is approximately distributed under the true likelihood, in place of exact simulation.
\item In \cite{friel4}, the Author proposes an efficient way to estimate the evidence based on the identity
\begin{equation}
{\pi}(y)=\frac{q(y\mid\param^{*})}{{z}(\param^{*})}\frac{p(\param^{*})}{{\pi}(\param^{*}\mid y)}\,,
\label{eq:chib_approx}
\end{equation}
which holds for all $\param^*\in\Theta$. Of course ${z}(\param^{*})$ and ${\pi}(\param^{*}\mid y)$ are unknown. \cite{friel4} devises the "population" version of the exchange algorithm that yields draws, $\{\param^{(i)}\}$, from the posterior distribution by transitioning from the prior (see Supplementary material). Additional auxiliary draws at each iteration of the MCMC scheme with the TNT sampler are used to give an estimate of the intractable normalising constant, $\hat{z}(\theta^{(i)})$, at each of these points. A kernel density approximation of the posterior, $\widehat{\pi}(\param\mid y)$, is also found using these points. The estimator of $\pi(y)$ is obtained by averaging different estimated values of $\pi(y)$ replacing the unknown quantities in \eqref{eq:chib_approx} by the estimated ones $\hat{z}(\theta^{(i)})$ and $\widehat{\pi}(\param\mid y)$, for a number of draws of the posterior that are close to posterior mean.
\end{itemize}

%%%%%%%%%%%%%%%%%%%%%%%%%%%%%%%%%%%%%%%%%%

Both applications to social networks involve comparisons against the Auto-RJ and the population exchange algorithms. The main weakness of the population exchange method is that it relies on (i) extensive simulations from the likelihood, increasing the computational burden and (ii) kernel density estimation of the target posterior distribution, which makes it impractical for use when the parameter space is high-dimensional. 
The application to the adolescent friendship network serves as an example where model comparisons based on unadjusted pseudolikelihoods are misleading, as opposed to comparisons based on fully adjusted pseudolikelihoods.

\subsection{Gaussian random walk updates}\label{section:updates}
The low-dimensional parameter space of the models considered in this paper allow to sample the target posterior distribution in one block. With the exemption of the auto-reversible jump exchange algorithm, all the other methods considered in this paper employed a random walk Metropolis strategy with a multivariate Gaussian proposal distribution.

A proposal distribution with a variance-covariance matrix in the form $\Sigma_{\lambda} = \lambda^2 I_d(B_0 + C^{-1})^{-1}$ was assumed \citep{chib2,mcmcpack}, to account for possible correlations between the model parameters. Here, $\lambda\in\rset^{+}$ denotes the Metropolis tuning scalar parameter and $B_0$ is the prior precision.
There are two options regarding the choice of the precision matrix $C^{-1}$: (i) it is the same as the negative Hessian $-\mathcal{H}_{\text{PL}}(\htheta_{MPLE})=-\nabla^{2}_{\param}\log{f^{}_{\text{PL}}(y\mid\param)}|_{\htheta_{MPLE}}$ when inference is based on the unadjusted pseudolikelihood or (ii) it is the same as the negative Hessian $-\nabla^{2}_{\param}\log{f(y\mid\param)}|_{\htheta_{MLE}}$ when inference is based on the adjusted pseudolikelihood.

The adjusted pseudo-posterior MCMC algorithm and the population exchange algorithm were tuned using the sample covariance matrix of the likelihood at the MLE from option (ii).
Here $B_0$ plays the role of regularising the Hessian. This strategy assumes that the Hessian is invertible. Obviously, a bad choice of the covariance matrix $\Sigma_{\lambda}$ will have an effect on posterior inference; a short MCMC run helped us decide on a value of the Metropolis tuning parameter in order to reach a reasonable mixing rate of around 25\% \citep{rosenthal}.

As regards the sampling algorithms based on tempered likelihoods, the parameters at all temperatures were updated jointly with Gaussian random walk proposals. It would be desirable to scale those proposals within different temperatures; it is appropriate to have wider proposals at lower temperatures so that the algorithm can explore the posterior support more effectively.
When updating the parameter vector $\theta_j$ in temperature $t_j$, \cite{wyse_review} chose a proposal from a Gaussian distribution which was centered at $\theta_j$ and with standard deviation $(t^{\alpha}\tau_p)^{-1/2}$, where $\tau_p$ is the proposal precision at temperature $t=1$. For their logistic regression example they chose a value of $\alpha$ such that the variability of the proposal near zero temperature would equal that of the prior. In the case of power posteriors, this was $\alpha = \log(\tau/\tau_p)/\log(t_{1})$.
This scaling led to reasonable acceptance rates in our experiments.

\subsection{Karate club network}\label{section:karate}
\begin{figure}[H]
\centering
\includegraphics[clip,trim={0 0 0 2cm},width=8cm,height=8cm]{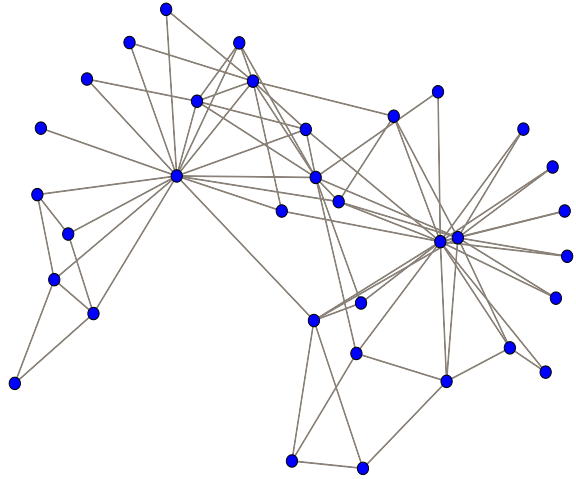}
\vspace{-2.6em}
\caption{Zachary's Karate Club graph.}
\label{fig:karate_club_graph}
\end{figure}

The Zachary's Karate Club network data, displayed in Figure \ref{fig:karate_club_graph}, represents a social network of friendships between 34 members of a karate club at a US university in the 1970. Three competing models were proposed by \cite{caimo2} to fit the data in the presence of degeneracy:
\begin{center}
\begin{tabular}{ll}
$\models_1$: & $q_1(y\mid \param_1)=\exp\big\{\param_{11}s_1(y)+\param_{12}v(y,\phi_v)\big\}$ \\
$\models_2$: & $q_2(y\mid \param_2)=\exp\big\{\param_{21}s_1(y)+\param_{22}u(y,\phi_u)\big\}$ \\
$\models_3$: & $q_3(y\mid \param_3)=\exp\big\{\param_{31}s_1(y)+\param_{32}v(y,\phi_v)+\param_{33}u(y,\phi_u)\big\}$, \\
\end{tabular}
\end{center}
where $s_1(y)=\sum_{i<j}^{}y_{ij}$ is the number of edges. The other model terms are defined below.
%%%%%%%%%%%%%%%%%%%%%%%%%%%%%%%%%%%%%%%%%%
\vspace{-1.1em}
%%%%%%%%%%%%%%%%%%%%%%%%%%%%%%%%%%%%%%%%%%
\textbf{Shared Partnership:} Let $EP_k(y)$, called the edgewise shared partnership statistic, denote the number of connected pairs with exactly k common neighbours. $EP_k(y)$ is a function of the triangle counts and as such, it is equivalent to modeling the high-order transitivities. The distribution of edgewise shared partnership can be modeled as a function of a single parameter by placing decreasing weights on the higher transitivities, leading to the geometrically weighted edgewise shared partnership (GWESP) statistic. GWESP is defined by:
%%%%%%%%%%%%%%%%%%%%%%%%%%%%%%%%%%%%%%%%%%
\begin{equation*}
v(y,\phi_v) = e^{\phi_v} \sum_{k=1}^{n-2}\left\{ 1-\left(1 - e^{-\phi_v} \right)^{k}\right\}EP_k(y)\,.
\end{equation*}
%%%%%%%%%%%%%%%%%%%%%%%%%%%%%%%%%%%%%%%%%%
\textbf{Geometrically Weighted Degree:} Let the degree count, $D_k(y)$, denote the number of pairs that have exactly k common neighbours. The number of stars is a function of the degrees, therefore $D_k(y)$ is equivalent to modeling the k-star statistic. The geometrically weighted degree (GWD) statistic enables to model all degree distributions as a function of single parameter by placing decreasing weights on the higher degrees. GWD is defined by:
%%%%%%%%%%%%%%%%%%%%%%%%%%%%%%%%%%%%%%%%%%
\begin{equation*}
u(y,\phi_u) = e^{\phi_u} \sum_{k=1}^{n-1}\left\{ 1-\left(1 - e^{-\phi_u} \right)^{k}\right\}D_k(y)\,.
\end{equation*}
%%%%%%%%%%%%%%%%%%%%%%%%%%%%%%%%%%%%%%%%%%
The scale parameters $(\phi_v,\phi_u)$ specify the decreasing rates of weights placed on the higher order terms, are treated as constants and are set to $(\phi_{v}\,,\phi_{u}) = (0.2\,,0.8)$. The main focus of this example will lie on the comparison between $\models_1$ and Model $\models_3$. Table \ref{tab:averageBF_comparison_karate_graph} makes clear that there is positive evidence in favor of $\models_1$ over Model $\models_3$, implying that the effect captured by the geometrically weighted degree network statistic does not enhance the observed network.

Thirty independent MCMC experiments with the Auto-RJ exchange algorithm were run, where each simulation consisted of 500,000 iterations, discarding the initial 50,000 as part of the burn-in. The auxiliary chain consisted of 300,000 iterations. Following arguments presented in \cite{Everitt}, the approximate exchange algorithm \citep{caimo1} that the Auto-RJ is using converges to the target distribution as the number of auxiliary draws tends to infinity.
There are 561 possible ties in the Karate Club network and so 300,000 auxiliary iterations should ensure that the long auxiliary MCMC run from the TNT sampler returns a draw that is approximately distributed under the true likelihood.

The Auto-RJ experiments yielded an average value of $BF^{}_{13}=13.177$ and between-model acceptance rate $\tilde{B}=5.5\%$. The low acceptance rate for the jump proposals suggests that the proposal distributions are not a good fit to each posterior model. As a result, the chain is mixing poorly and each model is not visited with the correct frequency. This, in turn, has an effect on the Bayes factor estimate.

The population exchange algorithm was implemented with 50 chains for 10,000 overall iterations after a burn-in period of 1,000 iterations, where 5,000 auxiliary iterations were used to generate an approximate draw of a network from the likelihood. A further 500 draws were used for the importance sampling estimate of the ratio of normalising constants between successive temperatures. We considered a temperature schedule $t_i=(i/50)^5,\,i=0,\ldots 50$. The closest 500 MCMC draws to the posterior mean of $\param$ were used to estimate $\pi(y)$.

Chib \& Jeliazkov's method drew a sample of 1,000,000 from the (corrected) pseudo-posterior distribution for estimating the evidence. The corresponding MCMC sampler was tuned appropriately, following the guidelines at Section \ref{section:updates} to obtain an overall acceptance rate of 20-25\%.

For the tempering schemes we used a temperature ladder with 101 rungs, $t_i=(i/100)^5,\,i=0,\ldots 100$. Within each temperature $t_i$, 30,000 samples were collected from the corresponding stationary distribution, after a burn-in period of 5,000 iterations.
This extended run increased considerably the computational expense but aided towards the reduction of estimation bias (see \cite{friel6} for general recommendations on the number of rungs and the length of the MCMC run at each temperature $t_i$). We note that the additional computation related to the control variates is a negligible fraction of the total computational cost.

The importance sampling algorithm for estimating the intractable normalising constant was carried out using 500 path points. For each point the TNT sampler was run for 5,000 iterations, followed by an extra 75,000 iterations thinned by a factor of 50, yielding 1500 networks. The adjustment algorithm took 3 min for each of the two models (Table \ref{tab:karate_graph_adj_times}). The overall time spent on the adjustments of the pseudolikelihood is accounted for in Table \ref{tab:averageBF_comparison_karate_graph}.
%%%%%%%%%%%%%%%%%%%%%%%%%%%%%%%%%%%%%%%%%%
\begin{table}[H]
\caption{Zachary karate club - CPU time in minutes for each adjustment phase.}
\vspace{-0.8em}
\centering
\begin{tabular}{lrrrr}
\toprule
Model & Mode & Curvature & Magnitude & Total\\
\hline
$\models_1$ & 0.336 & 0.002 & 2.688 & 3.026 \\
$\models_3$ & 0.417 & 0.002 & 2.874 & 3.293 \\
\bottomrule
\end{tabular}
\label{tab:karate_graph_adj_times}
\end{table}
%%%%%%%%%%%%%%%%%%%%%%%%%%%%%%%%%%%%%%%%%%
Results, shown in Table \ref{tab:averageBF_comparison_karate_graph}, demonstrate that the magnitude and direction of the Bayes factor estimates with the adjusted pseudolikelihood agree with the Auto-RJ results. In fact, working with the unadjusted pseudolikelhood in this example gives Bayes factor estimates of similar order of magnitude, but it is not recommended to work with such an approximation as we cannot be sure about its quality beforehand for a given dataset.

Additionally, there is good agreement with the evidence estimates from the population exchange MCMC run. Our procedure gets very accurate estimates of the true Bayes factor, but in a fraction of the time, which renders it a more appealing option.
%%%%%%%%%%%%%%%%%%%%%%%%%%%%%%%%%%%%%%%%%%
\begin{table}[H]
\caption{Zachary karate club - Average and standard deviation values of log-marginal likelihood, Bayes factor estimates and CPU time in minutes from thirty independent experiments, based on the unadjusted and the fully adjusted pseudolikelihood function. The CPU time corresponds to the total computational time required to apply an algorithm to both models of interest.}
\vspace{-0.8em}
\small
\centering
\setlength{\tabcolsep}{0.42em}
\begin{tabular}{llrrrr}
\toprule
Adjustment & Method &$\log\pi(y\mid \models_1)$ & $\log\pi(y\mid \models_3)$ &$BF^{}_{13}$ & CPU\\
\hline										
(a) No &Chib \& Jeliazkov     & -217.197 (0.01)& -219.842 (0.01) &14.088 (0.04) &14.20\\
&Stepping stones       & -216.805 (0.23)& -219.520 (0.26) &15.809 (4.71) &46.17 \\
&Power posteriors - TI & -216.798 (0.24)& -219.523 (0.27) &16.041 (5.01) &46.17\\
&Power posteriors - CTI& -217.077 (0.05)& -219.726 (0.04) &14.173 (0.81) &46.17 \\
\hline										
(b) Yes &Chib \& Jeliazkov  &-219.007 (0.01) &-221.766 (0.01) &15.776 (0.06)& 20.79 \\
&Stepping stones       & -218.765 (0.15) & -221.524 (0.20) & 16.192 (3.79) & 57.90 \\
&Power posteriors - TI & -218.763 (0.16) & -221.525 (0.21) & 16.302 (4.00) & 57.90 \\
&Power posteriors - CTI& -218.967 (0.02) & -221.716 (0.02) & 15.631 (0.42) & 57.90\\
\hline
&Population exchange   & -218.954 (0.02)& -221.703 (0.02) & 15.629 (0.35) & 1749.60\\
\hline
&Auto-RJ               &               -&                -& 13.177 (0.24) & 4777.86\\
\bottomrule
\end{tabular}
\label{tab:averageBF_comparison_karate_graph}
\end{table}
%%%%%%%%%%%%%%%%%%%%%%%%%%%%%%%%%%%%%%%%%%
Chib and Jeliazkov's one-block approach comes at a lower computational cost compared to TI. The Power posterior-related estimates are based on the improved trapezium rule (S.4). The potential gains made when estimating the evidence using power posteriors by correcting the numerical integration error have been illustrated previously \citep{friel6}. As regards the controlled thermodynamic integral estimate, the results are additionally based on a zero variance (ZV) control variates polynomial $P(\param)$ of dimension 2 (see Supplementary material).
%%%%%%%%%%%%%%%%%%%%%%%%%%%%%%%%%%%%%%%%%%
\begin{figure}[H]
\centering
\includegraphics[width=16cm,height=11cm]{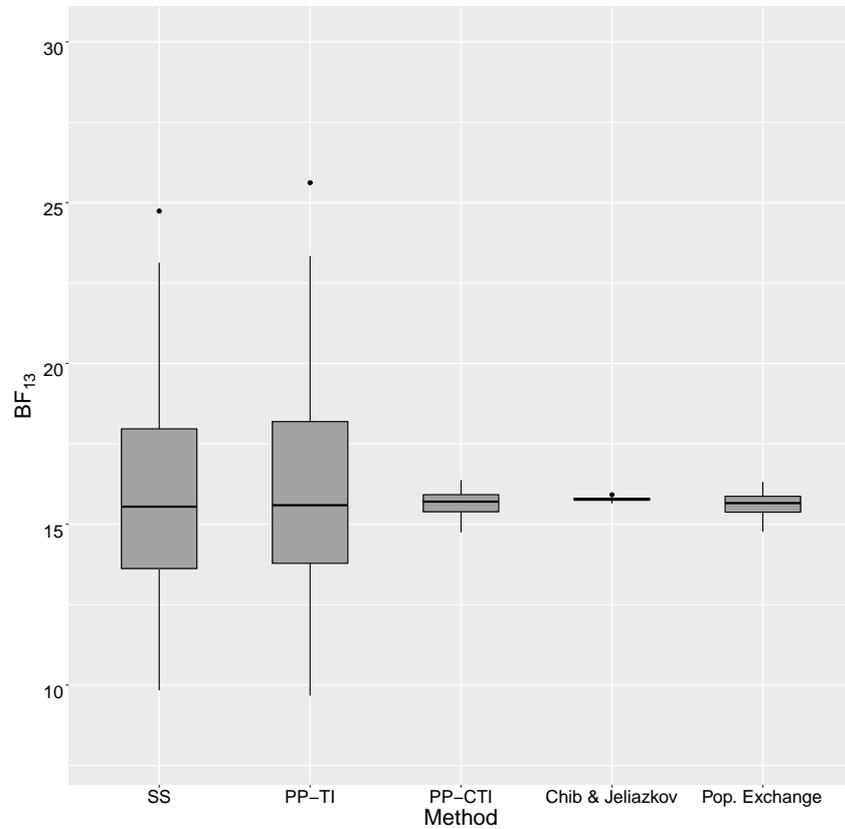}
\caption{Zachary karate club - Boxplots of estimated Bayes factors over thirty independent experiments. SS denotes the Stepping stones approach. PP-TI refers to the power posterior approach under the second-order quadrature method and PP-CTI refers to the power posterior approach under the second-order quadrature method and a controlled thermodynamic integral with a zero variance control variates polynomial $P(\param)$ of dimension 2.}
\label{fig:karate_club_BF_boxplots2}
\end{figure}
%%%%%%%%%%%%%%%%%%%%%%%%%%%%%%%%%%%%%%%%%%
There is very small discrepancy between the Stepping stone estimators and the Power posterior estimators based on second-order quadrature under the long temperature ladder. CTI (degree 2) achieves a massive variance reduction in the estimator variance. We see that this variance reduction transfers to estimates of the Bayes factor themselves, where the standard deviation of the CTI estimators (degree 2) is approximately $10\times$ lower compared to estimators based on TI.
Figure \ref{fig:karate_club_BF_boxplots2} displays that the CTI estimators (degree 2) are close to the Chib and Jeliazkov's estimators.

\subsection{Teenage Friends and Lifestyle Study}
\begin{figure}[H]
\centering
\includegraphics[clip,width=15cm,height=11cm]{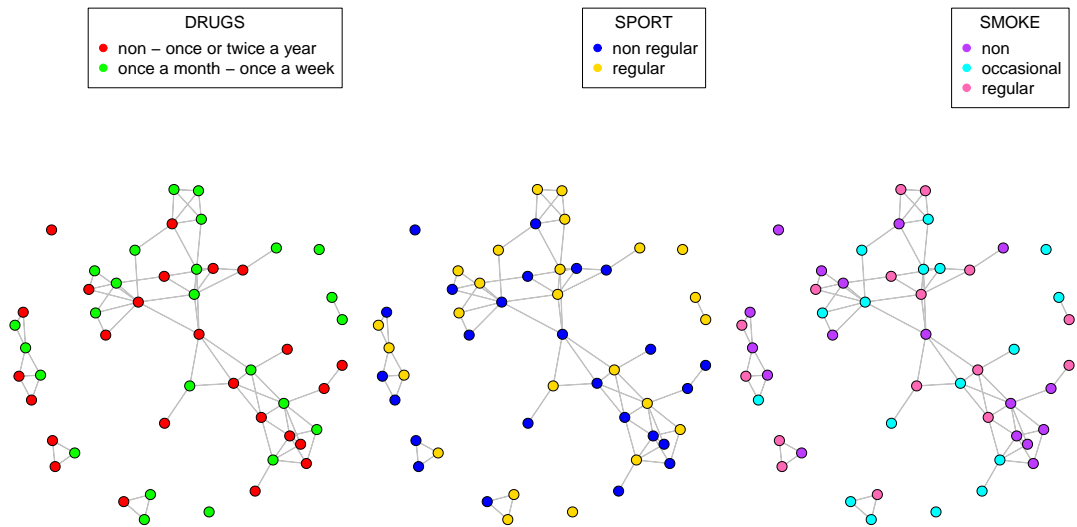}
\vspace{-6.8em}
\caption{50 girls from the Teenage Friends and Lifestyle Study dataset.}
\label{fig:friends_graph}
\end{figure}

The adolescent friendship network is a subset of a friendship network collected in the "Teenage Friends and Lifestyle Study" \citep{pearson}. The study records a network of friendships and substance use for a cohort of students in a secondary school in Glasgow, Scotland. Here we used an excerpt of 50 adolescent girls; the resulting network, displayed in Figure \ref{fig:friends_graph}, consists of 50 nodes and 39 edges. There are four covariates associated with each node: \textit{Drug usage}, \textit{Smoking status}, \textit{Alcohol usage} and \textit{Sport activity}.

Following \cite{caimo2}, we restrict our attention to three node covariates: \textit{Drug usage}, \textit{Sport activity}, and \textit{Smoking status}. We focus on the transitivity effect (expressed by the GWESP statistics with $\phi_{v} =\log{2}$), the degree heterogeneity (expressed by the GWD statistics with $\phi_u =0.8$) and the relationship between drug consumption and smoking (denoted by $s_4(y,x)=\textit{nodematch(c("smoke","drugs"))}$).

The homophily effect counts the number of edges for which two nodes share the same covariate value. When multiple relationships are studied, the "nodematch" statistic \citep{morris} counts only those on which all the covariate values match.  We compare two models:
\begin{center}
\begin{tabular}{ll}
$\models_1$: & $q_1(y\mid \param_1)=\exp\big\{\param_{11}s_1(y)+\param_{12}v(y,\phi_v)+\param_{13}u(y,\phi_u)\big\}$ \\
$\models_2$: & $q_2(y\mid \param_2)=\exp\big\{\param_{21}s_1(y)+\param_{22}v(y,\phi_v)+\param_{23}u(y,\phi_u)+\param_{24}s_4(y,x)\big\}$, \\
\end{tabular}
\end{center}
where $s_1(y)=\sum_{i<j}^{}y_{ij}$ is the number of edges.

Since the network is sparse, \cite{caimo2} incorporated this prior knowledge by setting the parameter value for the edges statistic equal to $-1$ and set the Multivariate Gaussian prior covariance matrix of each model to be a diagonal matrix with every entry equal to 5.

The Auto-RJ was again used as a reference. The simulation was run for 400,000 iterations (equating to 50 hours of CPU time), discarding the initial 50,000 as part of the burn-in, and the auxiliary chain consisted of 200,000 iterations.
The average Bayes factor (and standard deviation) based on thirty independent MCMC experiments was $BF^{}_{21}=1.186~(0.015)$.

This value shows slightly positive evidence in favor of $\models_2$ (but is rather close to one), revealing that the transitivity effect and the geometrically weighted degree distribution can explain the complexity of the observed network data, being important features of the graph and that the observed network is enhanced by the homophily effect of drug usage and smoking.
Similarly to Section \ref{section:karate}, the Auto-RJ suffers from low acceptance rates for the jump proposals: the between-model acceptance rate was equal to $\tilde{B}=8.2\%$.
%%%%%%%%%%%%%%%%%%%%%%%%%%%%%%%%%%%%%%%%%%
\vspace{-1.4em}
%%%%%%%%%%%%%%%%%%%%%%%%%%%%%%%%%%%%%%%%%%
For the population exchange algorithm we considered a temperature schedule $t_i=(i/100)^5,\,i=0,\ldots 100$, with 10,000 overall iterations after a burn-in period of 1,000 iterations. Each simulation consisted of an auxiliary MCMC run of length 5,000 as a proxy for an exact
sampler from the likelihood and a further 500 draws were used for the importance sampling estimate of the ratio of normalising constants between successive temperatures. The closest 500 MCMC draws to the posterior mean of $\param$ were used to estimate $\pi(y)$. This setting gave results which agree with our method, but came at a high computational cost (Table \ref{tab:averageBF_comparison_teenage_friends}).

In terms of estimating the intractable normalising constant for each model based on importance sampling in our method, we used the same setting as in Section \ref{section:karate}. In this example the adjustment algorithm took 5.5 min for each of the two models (Table \ref{tab:friends_graph_adj_times}). The overall time spent on the adjustments of the pseudolikelihood is accounted for in Table \ref{tab:averageBF_comparison_teenage_friends}.
%%%%%%%%%%%%%%%%%%%%%%%%%%%%%%%%%%%%%%%%%%
\begin{table}[H]
\caption{Teenage Friends and Lifestyle - CPU time in minutes for each adjustment phase.}
\vspace{-0.8em}
\centering
\begin{tabular}{lrrrr}
\toprule
Model & Mode & Curvature & Magnitude & Total\\
\hline
$\models_1$ & 0.291 & 0.002 & 5.204 & 5.497 \\
$\models_2$ & 0.297 & 0.002 & 5.094 & 5.393 \\
\bottomrule
\end{tabular}
\label{tab:friends_graph_adj_times}
\end{table}
%%%%%%%%%%%%%%%%%%%%%%%%%%%%%%%%%%%%%%%%%%
With the Chib \& Jeliazkov's method 1,000,000 MCMC updates from the (corrected) pseudo-posterior distribution were run. Regarding the TI-related schemes, we ran 30,000 Metropolis-Hastings updates for each temperature, after a burn-in period of 5,000 iterations,
while the tempering scale was partitioned as $t_i=(i/100)^5,\,i=0,\ldots 100$. In total, 3,030,000 iterations were used to estimate the evidence for each model.

For this example we observe that the Bayes factor estimates based on the adjusted pseudolikelihood agree with the Auto-RJ results (Table \ref{tab:averageBF_comparison_teenage_friends}). There is also a good agreement between the Stepping stone estimators and the Power posterior estimators before applying the ZV scheme.
In general, all estimators exhibit internal consistency; a 20-fold reduction in the evidence estimator variance is observed when applying the ZV control variates scheme.
%%%%%%%%%%%%%%%%%%%%%%%%%%%%%%%%%%%%%%%%%%
\begin{table}[H]
\caption{Teenage Friends and Lifestyle - Average and standard deviation values of log-marginal likelihood, Bayes factor estimates and CPU time in minutes from thirty independent experiments, based on the unadjusted and the fully adjusted pseudolikelihood function. The CPU time corresponds to the total computational time required to apply an algorithm to both models of interest.}
\vspace{-0.8em}
\small
\centering
\setlength{\tabcolsep}{0.42em}
\begin{tabular}{llrrrr}
\toprule
Adjustment & Method   &$\log\pi(y\mid \models_1)$ & $\log\pi(y\mid \models_2)$ &$BF^{}_{21}$ & CPU\\
\hline										
(a) No &Chib \& Jeliazkov     & -152.010 (0.01) &-153.153 (0.01) &0.319 (0.01) & 18.75\\
&Stepping stones       & -152.204 (0.14) &-153.316 (0.20) &0.340 (0.09) & 52.41\\
&Power posteriors - TI & -152.206 (0.15) &-153.312 (0.21) &0.343 (0.09) & 52.41\\
&Power posteriors - CTI& -152.215 (0.01) &-153.362 (0.01) &0.318 (0.01) & 52.41\\
\hline
(b) Yes &Chib \& Jeliazkov  &-233.740 (0.01) &-233.621 (0.01) &1.127 (0.01) & 28.13 \\
&Stepping stones       & -233.965 (0.12) & -233.801 (0.20) & 1.212 (0.28) & 67.44\\
&Power posteriors - TI & -233.967 (0.12) & -233.797 (0.20) & 1.221 (0.29) & 67.44\\
&Power posteriors - CTI& -233.975 (0.01) & -233.847 (0.01) & 1.137 (0.02) & 67.44\\
\hline
&Population exchange   & -233.934 (0.01)& -233.807 (0.01)& 1.135 (0.02) &7026.06\\
\hline
&Auto-RJ               & -& -& 1.186 (0.02) & 2977.08\\
\bottomrule
\end{tabular}
\label{tab:averageBF_comparison_teenage_friends}
\end{table}
%%%%%%%%%%%%%%%%%%%%%%%%%%%%%%%%%%%%%%%%%%
The Teenage Friends and Lifestyle Study network offers a scenario where any model comparisons based on the unadjusted pseudolikelihood are quite misleading, as the Bayes factor based on the unadjusted pseudolikelihood is $<1$, while the Bayes factor based on the adjusted pseudolikelihood is $>1$. There is a striking difference between the unadjusted pseudolikelihood-based estimates of the evidence and those estimates based on the fully adjusted pseudolikelihood.
All Bayes factor estimates slightly favour $\models_1$, which is in contrast with the Auto-RJ results and the results based on fully adjusted pseudolikelihoods.
Therefore, we can conclude that the benefits of conducting model selection for this network based on the unadjusted pseudolikelihood approximation are reduced. All in all, we highly recommend approximation of the evidence with the fully adjusted pseudolikelihood, which comes at a negligible computational cost.

%====================================================================================
%====================================================================================
\section{Discussion}\label{section:discussion}
In this paper we have presented a novel approach to marginal likelihood estimation of models with intractable normalising constants, which we applied to the challenging setting of exponential random graph models for social network analysis. We approximated a doubly-intractable posterior distribution with an intractable likelihood by a "standard" singly intractable posterior distribution with a tractable likelihood approximation. Our methodology is highly compatible with a plethora of evidence estimation techniques from the Bayesian toolbox.

Our experiments suggest that the one-block Metropolis-Hastings approach yields marginal likelihood estimators with low variability. It comes at a low computational cost and is suitable because it requires no further tuning of the MCMC algorithm and the low-dimensional parameter spaces in our experiments allow the parameter vector to be updated in a single block. For higher-dimensional MCMC problems, though, multi-block Metropolis-within-Gibbs updating strategies will be more suitable \citep{chib2}.
The Power posteriors algorithm can come at a higher computational cost, depending on the inverse temperature scheme. We suggest the use of the improved trapezoidal scheme of \cite{friel6} and the control variate technique \citep{oates} to vastly improve the statistical efficiency of the evidence estimate.
The Stepping stones estimators suffer from high variability.

We note that our approach avoids the heavy computational burden of repeated likelihood simulations, as in \cite{friel4} and \cite{johansen}. Here the likelihood simulations are performed only once for each competing model. The empirical results presented above suggest that our approach gives similar estimates of the Bayes factor to the computationally intensive population exchange algorithm, but at a fraction of the time. Simulation procedures for approximating a solution to the likelihood equation are more challenging when larger datasets are considered, but any simulation approach with increased dependence on likelihood simulations will be infeasible under these conditions. Our work should offer a more scalable approach.

Overall, the reasonable computational effort that needs to be put when working with our likelihood approximation extends the applicability of the proposed approach to other complex models like Gaussian Markov random fields and autologistic models; it will be particularly interesting to implement our proposed framework to large grids and networks with hundreds of nodes or models with many parameters.

% NOTE: AIS requires using temperatures!
%annealed importance sampling \citep{neal}

%====================================================================================
%====================================================================================
\bigskip
\begin{center}
{\large\bf SUPPLEMENTARY MATERIAL}
\end{center}

\begin{description}
\item[Evidence estimation strategies:] Detailed presentation of the strategies used in this paper for estimating the marginal likelihood, based on MCMC simulation. (.pdf file)
\item[R code:] R programs that can be used to replicate the Potts model study and exponential random graph example in Section 7 of this article. Please see the file README.txt contained within the accompanying zip file for more details.
\end{description}

%====================================================================================
%====================================================================================
%% References:
\bibliographystyle{Chicago}
\bibliography{main}

\begin{thebibliography}{}

\bibitem[\protect\citeauthoryear{Ardia, Ba\c{s}t\"{u}rk, Hoogerheide, and van
  Dijk}{Ardia et~al.}{2012}]{ardia_review}
Ardia, D., N.~Ba\c{s}t\"{u}rk, L.~Hoogerheide, and H.~van Dijk (2012).
\newblock {A comparative study of Monte Carlo methods for efficient evaluation
  of marginal likelihood}.
\newblock {\em Computational Statistics and Data Analysis\/}~{\em 56},
  3398--3414.

\bibitem[\protect\citeauthoryear{Besag}{Besag}{1972}]{besag3}
Besag, J. (1972).
\newblock {Nearest-neighbour systems and the auto-logistic model for binary
  data}.
\newblock {\em Journal of the Royal Statistical Society, Series B\/}~{\em
  34\/}(1), 75--83.

\bibitem[\protect\citeauthoryear{Besag}{Besag}{1975}]{besag1}
Besag, J. (1975).
\newblock {Statistical analysis of non-lattice data}.
\newblock {\em Statistician\/}~{\em 24}, 179--195.

\bibitem[\protect\citeauthoryear{Besag}{Besag}{1977}]{besag2}
Besag, J. (1977).
\newblock {Efficiency of pseudolikelihood estimation for simple Gaussian
  fields}.
\newblock {\em Biometrika\/}~{\em 64}, 616--618.

\bibitem[\protect\citeauthoryear{Bouranis, Friel, and Maire}{Bouranis
  et~al.}{2017}]{bouranismisp}
Bouranis, L., N.~Friel, and F.~Maire (2017).
\newblock {Efficient Bayesian inference for exponential random graph models by
  correcting the pseudo-posterior distribution}.
\newblock {\em Social Networks\/}~{\em 50}, 98--108.

\bibitem[\protect\citeauthoryear{Caimo and Friel}{Caimo and
  Friel}{2011}]{caimo1}
Caimo, A. and N.~Friel (2011).
\newblock {Bayesian inference for exponential random graph models}.
\newblock {\em Social Networks\/}~{\em 33}, 41--55.

\bibitem[\protect\citeauthoryear{Caimo and Friel}{Caimo and
  Friel}{2013}]{caimo3}
Caimo, A. and N.~Friel (2013).
\newblock {Bayesian model selection for exponential random graph models}.
\newblock {\em Social Networks\/}~{\em 35\/}(1), 11--24.

\bibitem[\protect\citeauthoryear{Caimo and Friel}{Caimo and
  Friel}{2014}]{caimo2}
Caimo, A. and N.~Friel (2014).
\newblock {Bergm: Bayesian exponential random graphs in R}.
\newblock {\em Journal of Statistical Software\/}~{\em 61\/}(2), 1--25.

\bibitem[\protect\citeauthoryear{Caimo and Mira}{Caimo and Mira}{2015}]{caimo4}
Caimo, A. and A.~Mira (2015).
\newblock {Efficient computational strategies for doubly intractable problems
  with applications to Bayesian social networks}.
\newblock {\em Statistics and Computing\/}~{\em 25\/}(1), 113--125.

\bibitem[\protect\citeauthoryear{Calderhead and Girolami}{Calderhead and
  Girolami}{2009}]{calderhead}
Calderhead, B. and M.~Girolami (2009).
\newblock {Estimating Bayes factors via thermodynamic integration and
  population MCMC}.
\newblock {\em Computational Statistics and Data Analysis\/}~{\em 53},
  4028–4045.

\bibitem[\protect\citeauthoryear{Chib}{Chib}{1995}]{chib1}
Chib, S. (1995).
\newblock {Marginal likelihood from the Gibbs output}.
\newblock {\em Journal of the American Statistical Association\/}~{\em
  90\/}(432), 1313--1321.

\bibitem[\protect\citeauthoryear{Chib and Jeliazkov}{Chib and
  Jeliazkov}{2001}]{chib2}
Chib, S. and I.~Jeliazkov (2001).
\newblock {Marginal likelihood from the Metropolis-Hastings output}.
\newblock {\em Journal of the American Statistical Association\/}~{\em 96},
  270--281.

\bibitem[\protect\citeauthoryear{Del~Moral, Doucet, and Jasra}{Del~Moral
  et~al.}{2006}]{delmoral}
Del~Moral, P., A.~Doucet, and A.~Jasra (2006).
\newblock {Sequential Monte Carlo Samplers}.
\newblock {\em Journal of the Royal Statistical Society, Series B\/}~{\em
  68\/}(3), 411--436.

\bibitem[\protect\citeauthoryear{Everitt, Johansen, Rowing, and Hogan}{Everitt
  et~al.}{2017}]{johansen}
Everitt, R., A.~Johansen, E.~Rowing, and M.~Hogan (2017).
\newblock {Bayesian model selection with un-normalised likelihoods}.
\newblock {\em Statistics and Computing\/}~{\em 27\/}(2), 403--422.

\bibitem[\protect\citeauthoryear{Everitt}{Everitt}{2012}]{Everitt}
Everitt, R.~G. (2012).
\newblock {Bayesian parameter estimation for latent Markov random fields and
  social networks}.
\newblock {\em {Journal of Computational and Graphical Statistics}\/}~{\em
  24\/}(4), 940--960.

\bibitem[\protect\citeauthoryear{Friel}{Friel}{2013}]{friel4}
Friel, N. (2013).
\newblock {Evidence and Bayes factor estimation for Gibbs random fields}.
\newblock {\em {Journal of Computational and Graphical Statistics}\/}~{\em
  22\/}(3), 518--532.

\bibitem[\protect\citeauthoryear{Friel, Hurn, and Wyse}{Friel
  et~al.}{2014}]{friel6}
Friel, N., M.~Hurn, and J.~Wyse (2014).
\newblock {Improving power posterior estimation of statistical evidence}.
\newblock {\em Statistics and Computing\/}~{\em 24}, 709--723.

\bibitem[\protect\citeauthoryear{Friel and Pettitt}{Friel and
  Pettitt}{2008}]{friel5}
Friel, N. and A.~N. Pettitt (2008).
\newblock {Marginal likelihood estimation via power posteriors}.
\newblock {\em Journal of the Royal Statistical Society, Series B\/}~{\em
  70\/}(3), 589--607.

\bibitem[\protect\citeauthoryear{Friel and Rue}{Friel and Rue}{2007}]{rue}
Friel, N. and H.~Rue (2007).
\newblock {Recursive computing and simulation-free inference for general
  factorizable models}.
\newblock {\em Biometrika\/}~{\em 93\/}(3), 661--672.

\bibitem[\protect\citeauthoryear{Friel and Wyse}{Friel and
  Wyse}{2012}]{wyse_review}
Friel, N. and J.~Wyse (2012).
\newblock {Estimating the evidence - a review}.
\newblock {\em Statistica Neerlandica\/}~{\em 66\/}(3), 288--308.

\bibitem[\protect\citeauthoryear{Gelman and Meng}{Gelman and
  Meng}{1998}]{gelman}
Gelman, A. and X.~Meng (1998).
\newblock {Simulating normalizing constants: from importance sampling to bridge
  sampling to path sampling}.
\newblock {\em Statistical Science\/}~{\em 13\/}(2), 163--185.

\bibitem[\protect\citeauthoryear{Geyer and Thompson}{Geyer and
  Thompson}{1992}]{geyer}
Geyer, C. and E.~Thompson (1992).
\newblock {Constrained Monte Carlo Maximum Likelihood for Dependent Data (with
  Discussion)}.
\newblock {\em Journal of the Royal Statistical Society, Series B\/}~{\em 54},
  657--99.

\bibitem[\protect\citeauthoryear{Hunter and Handcock}{Hunter and
  Handcock}{2006}]{hunter}
Hunter, D. and M.~Handcock (2006).
\newblock {Inference in curved exponential family models for networks}.
\newblock {\em Journal of Computational and Graphical Statistics\/}~{\em
  15\/}(3), 565--583.

\bibitem[\protect\citeauthoryear{Hunter, Handcock, Butts, Goodreau, {Morris},
  and {Martina}}{Hunter et~al.}{2008}]{ergm}
Hunter, D., M.~Handcock, C.~Butts, S.~Goodreau, {Morris}, and {Martina} (2008).
\newblock ergm: A package to fit, simulate and diagnose exponential-family
  models for networks.
\newblock {\em {Journal of Computational and Graphical Statistics}\/}~{\em
  24\/}(3), 1--29.

\bibitem[\protect\citeauthoryear{Kass and Raftery}{Kass and
  Raftery}{1995}]{kassr95}
Kass, R.~E. and A.~E. Raftery (1995).
\newblock {Bayes factors}.
\newblock {\em Journal of the American Statistical Association\/}~{\em
  90\/}(430), 773--795.

\bibitem[\protect\citeauthoryear{Koskinen}{Koskinen}{2004}]{koskinen}
Koskinen, J. (2004).
\newblock {Bayesian Analysis of exponential random graphs - Estimation of
  parameters and model selection}.
\newblock Technical Report~2, Department of Statistics, Stockholm University.

\bibitem[\protect\citeauthoryear{Lartillot and Phillipe}{Lartillot and
  Phillipe}{2006}]{lartillot}
Lartillot, N. and H.~Phillipe (2006).
\newblock {Computing Bayes factors using thermodynamic integration}.
\newblock {\em Systematic Biology\/}~{\em 55}, 195--207.

\bibitem[\protect\citeauthoryear{Lindsay}{Lindsay}{1988}]{lindsay}
Lindsay, B.~G. (1988).
\newblock {Composite likelihood methods}.
\newblock {\em {Contemporary Mathematics}\/}~{\em 80}, 221--239.

\bibitem[\protect\citeauthoryear{Liu}{Liu}{2001}]{liu}
Liu, J. (2001).
\newblock {\em {Monte Carlo strategies in scientific computing}}.
\newblock Springer Publishing Company, Incorporated.

\bibitem[\protect\citeauthoryear{Martin, Quinn, and Park}{Martin
  et~al.}{2011}]{mcmcpack}
Martin, A., K.~Quinn, and J.~Park (2011).
\newblock {{MCMCpack}: Markov chain Monte Carlo in {R}}.
\newblock {\em Journal of Statistical Software\/}~{\em 42\/}(9), 22.

\bibitem[\protect\citeauthoryear{Meng and Wong}{Meng and Wong}{1996}]{meng}
Meng, X. and W.~H. Wong (1996).
\newblock {Simulating ratios of normalizing constants via a simple identity: A
  theoretical exploration}.
\newblock {\em Statistica Sinica\/}~{\em 6}, 831--860.

\bibitem[\protect\citeauthoryear{M{\o}ller, Pettit, Reeves, and
  Bertheksen}{M{\o}ller et~al.}{2006}]{moller}
M{\o}ller, J., A.~Pettit, R.~Reeves, and K.~Bertheksen (2006).
\newblock {An efficient Markov chain Monte Carlo method for distributions with
  intractable normalising constants}.
\newblock {\em Biometrika\/}~{\em 93}, 451--458.

\bibitem[\protect\citeauthoryear{Morris, Handcock, and Hunter}{Morris
  et~al.}{2008}]{morris}
Morris, M., M.~Handcock, and D.~Hunter (2008).
\newblock {Specification of exponential-family random graph models: terms and
  computational aspects}.
\newblock {\em Journal of Statistical Software\/}~{\em 24\/}(4), 1--24.

\bibitem[\protect\citeauthoryear{Neal}{Neal}{2001}]{neal}
Neal, R. (2001).
\newblock {Annealed importance sampling}.
\newblock {\em Statistics and Computing\/}~{\em 11\/}(2), 125--139.

\bibitem[\protect\citeauthoryear{Oates, Papamarkou, and Girolami}{Oates
  et~al.}{2016}]{oates}
Oates, C., T.~Papamarkou, and M.~Girolami (2016).
\newblock {The controlled thermodynamic integral for Bayesian model evidence
  evaluation}.
\newblock {\em Journal of the American Statistical Association\/}~{\em
  111\/}(514), 634--645.

\bibitem[\protect\citeauthoryear{Pearson and Michell}{Pearson and
  Michell}{2000}]{pearson}
Pearson, M. and L.~Michell (2000).
\newblock {Smoke rings: social network analysis of friendship groups, smoking
  and drug-taking}.
\newblock {\em Drugs: Education, Prevention and Policy\/}~{\em 7\/}(1), 21--37.

\bibitem[\protect\citeauthoryear{Potts}{Potts}{1952}]{potts}
Potts, R.~B. (1952).
\newblock {Some generalized order-disorder transformations}.
\newblock {\em Proceedings of the Cambridge Philosophical Society\/}~{\em 48},
  106--109.

\bibitem[\protect\citeauthoryear{{R Core Team}}{{R Core Team}}{2017}]{rsoft}
{R Core Team} (2017).
\newblock {\em R: A Language and Environment for Statistical Computing}.
\newblock Vienna, Austria: R Foundation for Statistical Computing.

\bibitem[\protect\citeauthoryear{Reeves and Pettitt}{Reeves and
  Pettitt}{2004}]{reeves}
Reeves, R. and A.~N. Pettitt (2004).
\newblock {Efficient recursions for general factorisable models}.
\newblock {\em Biometrika\/}~{\em 91\/}(3), 751--757.

\bibitem[\protect\citeauthoryear{Ribatet, Cooley, and Davison}{Ribatet
  et~al.}{2012}]{ribatet}
Ribatet, M., D.~Cooley, and A.~Davison (2012).
\newblock {Bayesian inference from composite likelihoods, with an application
  to spatial extremes}.
\newblock {\em Statistica Sinica\/}~{\em 22}, 813--845.

\bibitem[\protect\citeauthoryear{Robins, Snijders, Wang, Handcock, and
  Pattison}{Robins et~al.}{2007}]{robins2}
Robins, G., T.~Snijders, P.~Wang, M.~Handcock, and P.~Pattison (2007).
\newblock {Recent developments in exponential random graph (p*) models for
  social networks}.
\newblock {\em Social Networks\/}~{\em 29}, 192--215.

\bibitem[\protect\citeauthoryear{Rosenthal and Roberts}{Rosenthal and
  Roberts}{2001}]{rosenthal}
Rosenthal, J. and G.~Roberts (2001).
\newblock {Optimal scaling for various Metropolis-Hastings algorithms}.
\newblock {\em Statistical Science\/}~{\em 16\/}(4), 351--367.

\bibitem[\protect\citeauthoryear{Snijders, Pattison, Robins, and
  Handcock}{Snijders et~al.}{2006}]{snijders_recent_2007}
Snijders, T., P.~Pattison, G.~Robins, and M.~Handcock (2006).
\newblock {New specifications for exponential random graph models}.
\newblock {\em Sociological Methodology\/}~{\em 36\/}(1), 99–153.

\bibitem[\protect\citeauthoryear{Stoehr and Friel}{Stoehr and
  Friel}{2015}]{friel1}
Stoehr, J. and N.~Friel (2015).
\newblock {Calibration of conditional composite likelihood for Bayesian
  inference on Gibbs random fields}.
\newblock In {\em AISTATS, Journal of Machine Learning Research: W \& CP},
  Volume~38, pp.\  921--929.

\bibitem[\protect\citeauthoryear{Stoehr, Pudlo, and Friel}{Stoehr
  et~al.}{2016}]{GiRaF}
Stoehr, J., P.~Pudlo, and N.~Friel (2016).
\newblock {\em {GiRaF: Gibbs random fields analysis}}.
\newblock R package version 1.0.

\bibitem[\protect\citeauthoryear{Strauss and Ikeda}{Strauss and
  Ikeda}{1990}]{strauss}
Strauss, D. and M.~Ikeda (1990).
\newblock {Pseudolikelihood estimation for social networks}.
\newblock {\em Journal of the American Statistical Association\/}~{\em 85},
  204--212.

\bibitem[\protect\citeauthoryear{Thiemichen, Friel, Caimo, and
  Kauermann}{Thiemichen et~al.}{2016}]{thiemichen}
Thiemichen, S., N.~Friel, A.~Caimo, and G.~Kauermann (2016).
\newblock {Bayesian exponential random graph models with nodal random effects}.
\newblock {\em Social Networks\/}~{\em 46}, 11--28.

\bibitem[\protect\citeauthoryear{Varin, Reid, and Firth}{Varin
  et~al.}{2011}]{varin1}
Varin, C., N.~Reid, and D.~Firth (2011).
\newblock {An overview of composite likelihood methods}.
\newblock {\em {Statistica Sinica}\/}~{\em 21\/}(1), 5--42.

\bibitem[\protect\citeauthoryear{Wang and Atchade}{Wang and
  Atchade}{2014}]{wang}
Wang, J. and Y.~F. Atchade (2014).
\newblock {Bayesian inference of exponential random graph models for large
  social networks}.
\newblock {\em Communications in Statistics - Simulation and
  Computation\/}~{\em 43}, 359--377.

\bibitem[\protect\citeauthoryear{Wasserman and Pattison}{Wasserman and
  Pattison}{1996}]{wasserman}
Wasserman, S. and P.~Pattison (1996).
\newblock {Logit models and logistic regression for social networks: I. An
  introduction to Markov graphs and p*}.
\newblock {\em Psycometrica\/}~{\em 61}, 401--425.

\bibitem[\protect\citeauthoryear{Xie, Lewis, Fan, Kuo, and Chen}{Xie
  et~al.}{2011}]{xie}
Xie, W., P.~O. Lewis, Y.~Fan, L.~Kuo, and M.~Chen (2011).
\newblock {Improving marginal likelihood estimation for Bayesian phylogenetic
  model selection}.
\newblock {\em Systematic Biology\/}~{\em 60\/}(2), 150--160.

\end{thebibliography}

\end{document}